# LIGHTWEIGHT LEARNING FOR GRANT-FREE ACTIVITY DETECTION IN CELL-FREE MASSIVE MIMO NETWORKS

ALI ELKESHAWY[1] (Student Member, IEEE), HAÏFA FARÈS [1] (Member, IEEE), AND AMOR NAFKHA[1] (Senior Member, IEEE)
[1]IETR - UMR CNRS 6164, CentraleSupélec, avenue de la Boulaie - CS 47601 35576, CESSON-SEVIGNE Cedex, France
Email: {ali-fekry-ali-hassan.elkeshawy, haifa.fares, amor.nafkha}@centralesupelec.fr

Corresponding author: ALI ELKESHAWY

**ABSTRACT** Grant-free random access (GF-RA) is a promising access technique for massive machine-type communications (mMTC) in future wireless networks, particularly in the context of 6G systems. Within the context of GF-RA, this study investigates the efficiency of employing supervised machine learning techniques to tackle the challenges on device activity detection (AD). GF-RA addresses scalability by employing non-orthogonal pilot sequences, which provides an efficient alternative comparing to conventional grant-based random access (GB-RA) technique that are constrained by the scarcity of orthogonal preamble resources. In this paper, we propose a novel lightweight data-driven algorithmic framework, represented by a single-layer perceptron (SLP), specifically designed for activity detection in GF-RA for mMTC in cell-free massive multiple-input multiple-output (CF-mMIMO) networks. We propose two distinct framework deployment strategies, centralized and decentralized, both tailored to streamline the proposed approach implementation across network infrastructures. Moreover, we introduce optimized post-detection methodologies complemented by a clustering stage as well as a post processing used to combine efficiently all collaborative decisions to enhance overall detection performance. Our 3GPP-compliant simulations demonstrate that the proposed algorithm achieves state-of-the-art performance in model-based activity detection, attaining a 99% detection probability with a low false alarm rate, while significantly reducing computational complexity. This highlights its practical effectiveness and suitability for real-world deployment.

**INDEX TERMS** 6G, Activity Detection, Grant-Free Random Access, Cell-Free massive MIMO Networks, Deep Learning, massive Machine-Type Communications, Sparsity.

## I. INTRODUCTION

An important step toward adopting internet of things (IoT) applications has been taken with the introduction of massive machine-type communications (mMTC) services [1]. This marks the beginning of a new era where devices, constrained by limited memory and energy availability, send small amounts of data sporadically or regularly at short intervals. The widespread proliferation of these devices in large areas presents significant challenges for current wireless networks, especially when it comes to managing the connectivity of such a vast number of connected devices. Within this context, Cell-free massive multiple-input multiple-output (CF-mMIMO) technology provides a promising solution to overcome mMTC challenges by significantly improving network coverage and supporting higher device densities, making it a key enabler for future wireless communication systems [2], [3].

Grant-based random access (GB-RA) techniques have been extensively investigated [4]–[6], and it is well established that each device must select a preamble from a limited set of orthogonal sequences to join the wireless network. However, since the number of orthogonal sequences is finite, caused by restricted coherence time and sequence length, there is an increased probability that two or more devices will choose the same sequence, resulting in collision and consequently, wireless access failure [7].

Thus, many grant-free (GF) access strategies have been proposed to solve the limitations of GB-RA techniques [8]. These strategies allow devices to join a wireless network without express permission. In GF scenarios, every device is assigned a unique non-orthogonal pilot sequence, avoiding the constraint of orthogonal sequence assignment because of the short channel coherence time. Therefore, GB-RA is seen as a feasible method to guarantee network connectivity by effectively controlling non-orthogonal pilot interference and reducing network access latency [9]. Accurate channel estimation is crucial for efficient data communication in mMTC networks, which requires the detection of active devices, a task heavily affected by the sparse activity patterns typical of machine-type devices. We are investigating techniques for identifying which devices are actively transmitting in a given time slot within a GF access system. Our focus is on sparse support recovery methods, which allow us to efficiently detect the small number of active devices among a larger pool of potential transmitters.

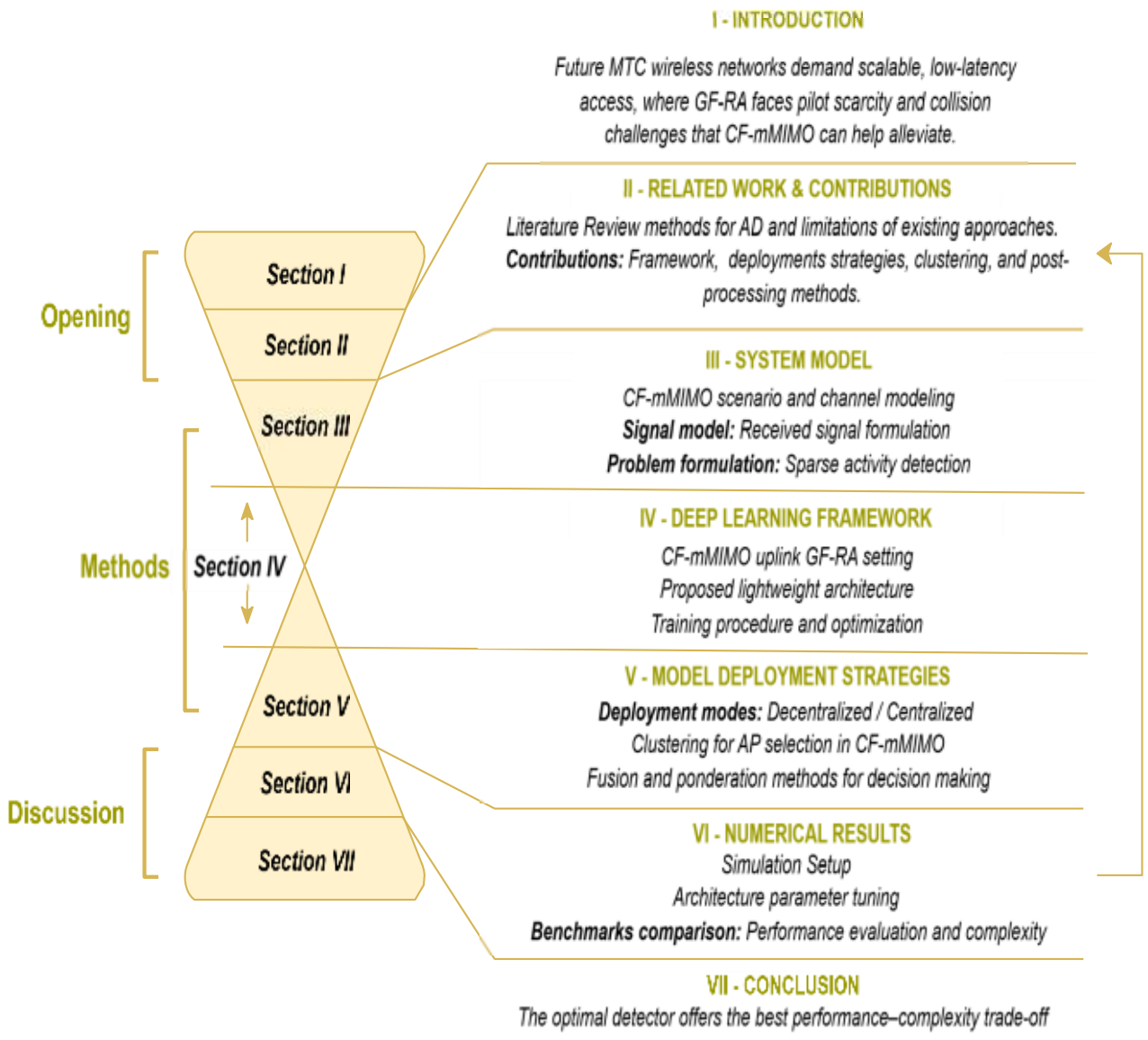

FIGURE 1. Manuscript Structure Overview.

Techniques for identifying sparse support can be categorized into three primary groups: compressive sensing (CS) methods [10]–[17], covariance-based (CV) approaches [18]–[21], and data-driven strategies [22], [23]. These methodologies represent distinct paradigms for addressing the challenge of sparse support recovery in various applications. Detecting active devices within grant-free random access for mMTC in CF-mMIMO networks, as shown in Fig. 2 (c), presents considerable difficulties, underscoring the necessity for sophisticated detection methods. GF access is a communication protocol that allows devices to transmit without prior scheduling permissions from the network controller, advantageous for environments requiring low latency and minimal signaling overhead, such as mMTC applications. In CF-mMIMO systems, as illustrated in Fig. 2(a), the stability provided by coherent blocks is critical to efficiently decode unscheduled transmissions from multiple devices enabled by GF access, thereby enhancing network performance and scalability.

The remainder of this paper is structured as follows: Section II presents related work in the field and highlights the main contributions of this paper. Section III introduces the system model. Section IV provides a detailed description of the DL framework, including the model architecture and its training phase. Section V provides the deployment strategies in the inference phase for the centralized or decentralized approach. Section VI presents the numerical results, demonstrating the effectiveness of the proposed algorithms. Finally, the paper is concluded in Section VII. [1]

TABLE 1. List of Acronyms.

| | |
|---|---|
| 3GPP | 3rd generation partnership project |
| 6G | Sixth-Generation |
| AD | Activity Detection |
| AMP | Approximate Message Passing |
| ANN | Artificial Neural Networks |
| APs | Access Points |
| CF-mMIMO | Cell-Free Massive Multiple-Input Multiple-Output |
| CNN | Convolutional Neural Network |
| CS | Compressive Sensing |
| CS-MUD | CS Multi-User Detection |
| CDF | Cumulative Distribution Function |
| CPU | Central Processing Unit |
| CV | Covariance-Based |
| DL | Deep Learning |
| DMLP | Deep Multilayer Perceptron |
| DNN | Deep Neural Network |
| FA | False Alarm |
| FISTA | Fast Iterative Shrinkage-Thresholding Algorithm |
| GB-RA | Grant-based Random Access |
| GF | Grant-Free |
| GF-RA | Grant-Free Random Access |
| GPU | Graphics Processing Unit |
| IoT | Internet of Things |
| ISTA | Iterative Soft Thresholding Algorithm |
| LASSO | Least Absolute Shrinkage and Selection Operator |
| LSF | Large-Scale Fading |
| MAC | Multiply-Accumulate |
| MIMO | Multiple-Input Multiple-Output |
| MD | Miss Detection |
| MMV | Multiple Measurement Vector |
| mMTC | Massive Machine-Type Communication |
| ML | Machine Learning |
| NN | Neural Network |
| ReLU | Rectified Linear Unit |
| ROC | Receiver Operating Characteristics |
| SLP | Single-Layer Perceptron |
| SSF | Small-Scale Fading |
| UMa-LoS | Urban Macro Line-of-Sight |
| ZC | Zadoff-Chu |

## II. RELATED WORKS

In this section, a concise overview of the literature on sparse support recovery relevant to GF-RA protocols is provided.

The device activity (AD) problem is formulated as a compressive sensing problem in [10], [11], with the goal of reconstructing a sparse signal from a restricted number of measurements. In GF-RA, these measurements correspond to the signal received at a given access point (AP), while the sparse signal represents the sporadic activity patterns of the MTC devices.

[1]See Fig. 1, which summarizes the overall structure of the manuscript for the reader.

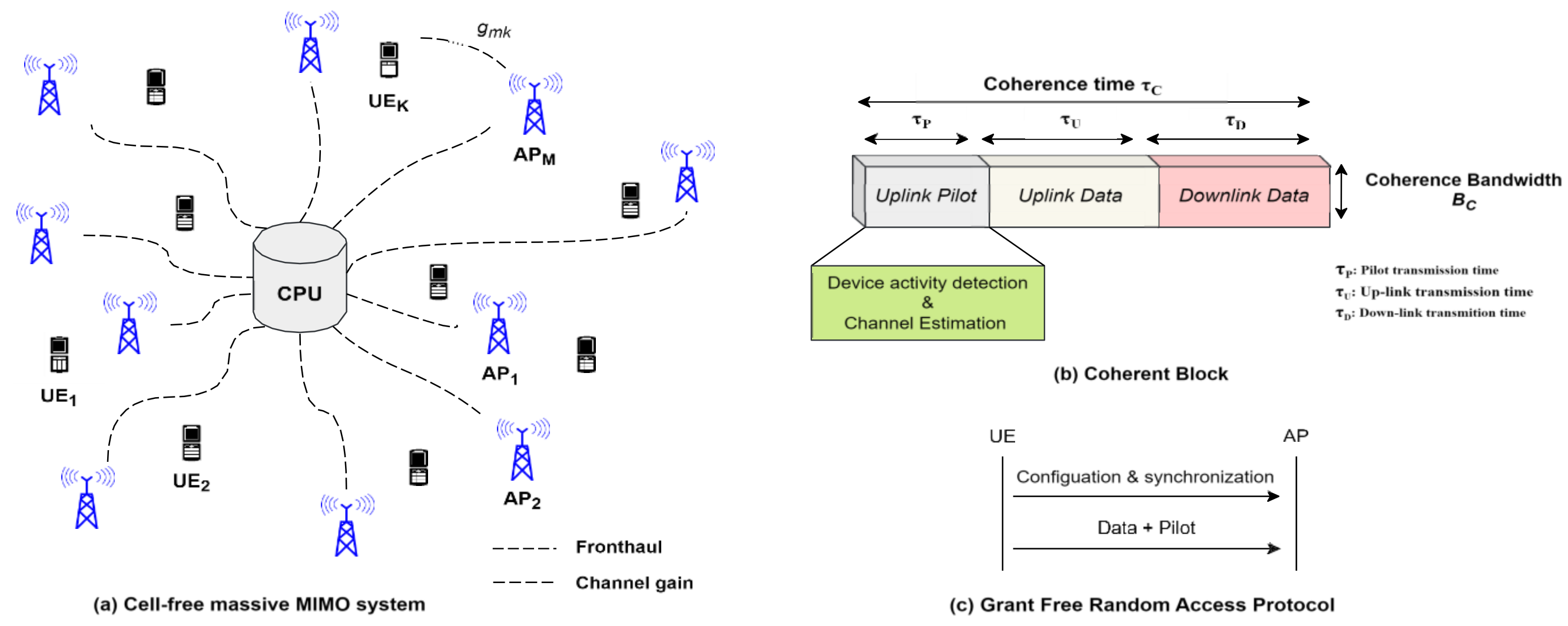

**FIGURE 2.** Cell-free network model for mMTC.

Several algorithms based on CS have been developed for joint channel estimation and device activity detection; among them, we can mention greedy pursuit algorithms which are widely used for sparse signal recovery [12], [13] and the approximate message passing (AMP) algorithm [14]. Additionally, the sparse inference problem in the compressive sensing context can be formulated as a least absolute shrinkage and selection operator (LASSO) optimization problem [24], which can be addressed using various algorithms, including convex relaxation methods such as the iterative soft thresholding algorithm (ISTA) and its accelerated version, FISTA (fast iterative shrinkage-thresholding algorithm) [25]. In [15], authors proposed a CS-based method to efficiently handle the massive number of devices in non-coherent mMIMO systems, that simultaneously transmit data without requiring dedicated resources or scheduling grants from the base station.

The CV-based method is presented in [18], showing better performance than current CS-based techniques. In fact, while CS techniques are well-suited for scenarios with moderate device activity and sufficient pilot resources, the covariance-based approach provides a more scalable and flexible solution for GF-RA in ultra-dense mMTC networks, particularly when the number of active devices exceeds the pilot sequence length. Moreover, in [19], authors suggest a CV-based method for joint activity and data detection, along with a performance analysis that underscores its superior advantages compared to other existing methods.

In [20], authors make several important contributions to the field of improving activity detection for mMTC in CF-mMIMO networks. They investigated GF-RA and modeled device activity detection as a maximum likelihood detection problem. Their approach, initially introduced in [21], focuses on device AD via a single dominating AP, marking a noteworthy advancement. In addition, authors propose a novel clustering-based activity detection technique that promises to enhance detection efficiency at the expense of higher computation complexity on the central processing unit (CPU). Their results highlight how CF-mMIMO networks can greatly improve activity identification when a large number of devices are connected, demonstrating the potential of this method in future wireless networks aiming to facilitate IoT services.

Authors in [26] propose a deep learning-based method for user activity detection in CF-mMIMO networks. Their method, which does not require prior large-scale fading coefficients, utilizes a fully grant-free approach, relying on non-orthogonal pilot sequences. Unlike traditional methods, no clustering approach is employed, and the activity detection model is based on signal projections from all access points in the network. They employ a convolutional neural network (CNN) to merge data from distributed antennas, outperforming covariance-based methods in detecting user activity. The proposed solution addresses challenges in large-scale fading and pilot contamination, demonstrating the effectiveness of data-driven methods in this context. Additionally, the model's performance is influenced by various network parameters, highlighting the importance of network configuration in achieving optimal activity detection.

To greatly improve device activity detection in mMTC scenarios under GF-RA protocols, authors in [22] suggest a dual-strategy method. Their work aims at robust and efficient AD by introducing two deep learning (DL) algorithms for sparse support recovery. They include a feature selection step, which focuses on selecting various pilot sequences with good cross-correlation properties to optimize the neural network (NN) design. The authors elaborate colocated (cellular) configuration and explore different kind of pilot sequences to enhance the activity detection performance. Their results show how effective zadoff-chu (ZC) sequences are as pilots, exhibiting higher activity detection probability with less computing load and providing a viable way to handle massive connectivity issues.

TABLE 2. Benchmark of Activity Detection Approaches: Strengths and Limitations

| Approach | Strengths | Limitations |
|---|---|---|
| **CS-based [12] [13] [24] [25]** | - Effective for sparse recovery<br>- Suitable for massive access | - Iterative and computationally expensive<br>- Requires parameter tuning |
| **Covariance-based Maximum likelihood [18] [21] [20]** | - Statistically grounded | - Requires large-scale fading coefficients<br>- Sensitive to channel uncertainty<br>- High computational load |
| **Deep Learning (CNN) [26]** | - LSF-free detection<br>- Learns spatial features | - Requires input from all AP signals<br>- Sensitive to AP count<br>- Projection-based detection<br>- High complexity |
| **Deep Learning [22]** | - Effective pilot type selection | - Cellular deployment only<br>- No clustering support |
| **Proposed Method** | - Lightweight and scalable<br>- LSF-independent detection<br>- No preprocessing required<br>- Simple computations<br>- Supports centralized and decentralized scenarios<br>- Non-iterative structure and low computational cost with respect to the CS-based approach<br>- Detection independent of LSF and robust to practical impairments with respect to the covariance-based approach<br>- Operates on cluster-level input, robust to low AP density, and signal-based, with respect to the CNN approach<br>- Supports clustering and decentralization with respect to the deep learning method | - Performance may depend on device density<br>- Requires fine-tuning for newly joined devices |

To overcome the computational complexity of CS multi-user detection (CS-MUD) in huge mMTC systems, in [23], authors present an innovative deep learning technique utilizing a specially designed block-restricted activation nonlinear unit, suitable to efficiently leverage the block sparse structure inherent in systems with many antennas or wide bandwidths. This approach enhances the performance of CS-MUD by offering a ten-fold reduction in computing time.

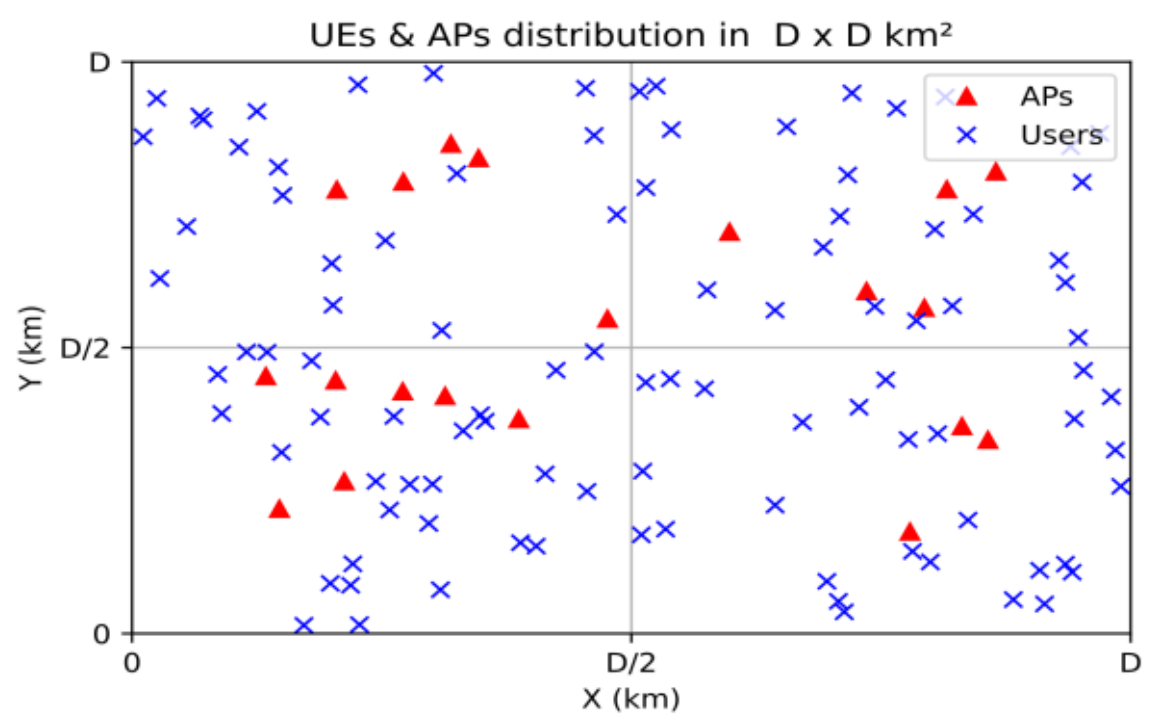

FIGURE 3. Distribution of devices and access points within the network.

Table 2 provides a comparative summary of state-of-the-art activity detection methods, outlining their advantages and limitations, and indicating the challenges addressed by the proposed approach.

### A. PAPER CONTRIBUTIONS

Motivated by the above discussion, in this paper, we propose a learning-based model to improve activity detection in grant-free random access for mMTC in CF-mMIMO network. Our main contributions can be resumed as follows:

- A robust learning-based model is proposed to allow accurate AD of the device in a GF-RA. This model leverages a single-layer perceptron (SLP) architecture to efficiently process symbols received during the random access slot, focusing on sparse support recovery for activity detection.
- Two deployment strategies have been developed for the proposed model: decentralized and centralized. In the decentralized approach, the model is distributed across all APs in the network, enabling localized and independent processing at each node. In contrast, the centralized approach consolidates data from all APs into the CPU, where computations are performed in a unified manner. While the two strategies differ in their operational workflows, they are built upon the same model architecture, ensuring consistency in performance and functionality.
- The architectural parameters of the proposed model are carefully analyzed through numerical evaluation to ensure optimal performance. This process aims to achieve a high detection probability in activity detection while minimizing the risk of overfitting, ensuring the model's reliability and generalization across diverse scenarios.
- Recognizing the computational complexities associated with involving all APs in device activity detection within a CF-mMIMO network, we propose a clustering method to improve efficiency and performance. The method is rigorously developed and evaluated under both decen-tralized and centralized strategies, addressing various implementation considerations. By carefully selecting a subset of APs with strong channel conditions for each device rather than engaging the entire network, the clustering approach optimizes the detection process. Additionally, this method is designed to operate independently of the core algorithm, effectively decoupling the clustering process and enhancing scalability.
- An extended analysis is conducted to evaluate the performance of various activity detection algorithms, including the proposed learning-based model and CV-based methods from the literature [25], within both CF-mMIMO and co-located mMIMO. These algorithms are rigorously tested under realistic scenarios, such as dense AP deployments, varying cell area sizes, and power-limited mMTC conditions. The results show that the proposed approach delivers improved performance compared to the methods outlined in [20], demonstrating its effectiveness in challenging network environments.

## III. SYSTEM MODEL

Consider a CF-mMIMO wireless network consisting of $M$ APs, each equipped with $N$ antennas, distributed across the coverage area and connected via front-haul connections to CPU. The APs serve a set of $K$ single-antenna MTC devices. Due to the sporadic nature of traffic in the mMTC massive access scenario, only a small fraction of the $K$ devices will be active at a given time instant.

In the present work, a grant-free random access (GF-RA) framework is considered, where devices transmit pilot sequences without prior scheduling or dedicated grants from the central unit. Each device makes autonomous transmission decisions based on a fixed activation probability, $\epsilon \ll 1$. The activity status of the $k$-th device is denoted by $a_k \in \{0, 1\}$, where $k \in \{1, 2, \ldots, K\}$ and $a_k = 1$ indicates that the $k$-th device is active, and $a_k = 0$ indicates that the $k$-th device is not active in the timeslot. The activation and inactivation probabilities of the $k$-th device are denoted by $\Pr(a_k = 1) = \epsilon$, and $\Pr(a_k = 0) = (1-\epsilon)$ respectively. The resulting vector $\mathbf{a} = (a_1, a_2, \ldots, a_K)^t$ represents the activity of the $K$ devices, and it is sparse vector due to the sporadic communication nature of mMTC. The set of active devices is represented by $\mathcal{A}$, where $\mathcal{A} = \{k \in 1, 2, \ldots, K : a_k = 1\}$. Additionally, it is noteworthy that the activity follows the Bernoulli distribution.

Considering the geographic distribution of the devices, we assume independent channel realizations between devices and access points. Moreover, we suppose that the $N$ antennas at each AP are sufficiently separated, so that each path provides an independent fading. The channel gain between $k$-th device and $n$-th antenna of the $m$-th access point is denoted by $g_{mk}^{(n)}$ and it can be modeled as follows:

$$g_{mk}^{(n)} = \beta_{mk}^{1/2} \times h_{mk}^{(n)}, \tag{1}$$

where $\beta_{mk}$ is defined as the large-scale fading (LSF) coefficient between the $m$-th access point and the $k$-th device as specified in 3GPP standard [27], and it is expressed as :

$$\begin{aligned} \beta_{mk} &= PL + F_{mk} \\ PL &= \begin{cases} PL_1, & 10m \leq d_{2D} \leq d'_{BP} \\ PL_2, & d'_{BP} < d_{2D} \leq 5km \end{cases} \\ PL_1 &= 28.0 + 22\log_{10}(d_{3D}) + 20\log_{10}(f_c) \\ PL_2 &= 28.0 + 40\log_{10}(d_{3D}) + 20\log_{10}(f_c) \\ &\quad - 9\log_{10}\left((d'_{BP})^2 + (h_{BS} - h_{UT})^2\right), \end{aligned} \tag{2}$$

The LSF coefficient $\beta_{mk}$ consists of two elements: the standard path loss model $PL$, pertinent to urban macro line-of-sight (UMa-LoS) conditions, and the shadow fading component modeled as a Gaussian random variable, $F_{mk} \sim \mathcal{N}(0, \sigma_{sh}^2)$. Shadowing may introduce correlation between antennas, especially if they are in close proximity or have similar line-of-sight (LoS) paths to a device. This correlation can affect the independence of fading paths in such scenarios. The path loss model is contingent on the horizontal distance $d_{2D}$, gauging the separation between each device $k$ and access pont $AP_m$, with $d'_{BP}$ denoting the break-point distance beyond which the model switches from model $PL_1$ to model $PL_2$, accounting for the height differences between devices and access points. Furthermore, the small-scale fading (SSF) coefficient is denoted by $h_{mk}^{(n)} \sim \mathcal{CN}(0, 1)$ and remains constant for some consecutive activity vectors, where the activity can change up to multiple times before the channel changes.

We assume that the large-scale fading (LSF) coefficients $\beta_{mk}$ are known at the CPU, as considered in [20], because they vary much more slowly compared to small-scale fading. Since LSF depends on relatively stable factors such as path loss and shadowing, it remains consistent over multiple coherence intervals, making it feasible to estimate in advance and use as prior information. Estimation techniques for the LSF coefficients are discussed in works such as [28], [29]. In the sequel of this paper, we consider a block fading scenario where each channel coefficient remains unchanged during a coherence time interval, and each channel realization is independent of all other realizations.

In the uplink of a narrow-band mMTC system within a square area of $D \times D$ km$^2$, assigning orthogonal pilot sequences to devices is unfeasible due to limited channel coherence time $\tau_C$ defined as the duration over which the channel remains approximately constant. The mMTC network typically supports a large number of devices, far exceeding the number of symbols $T_C$ that can be transmitted in the coherence block, as illustrated in Fig. 2(b), where $T_C$ is defined as $T_C = \tau_C \times B_C$, with $B_C$ representing the coherence bandwidth. The phenomenon of pilot contamination arises because each active device transmits non-orthogonal pilot sequences $\mathbf{s_k}$ during a random access slot. The sequences are drawn from a gaussian distribution, i.e., $\mathbf{s_k} \in \mathbb{C}^{L\times 1}$, where $L$ is the pilot sequence length. The length $L$ constitutes a specific reserved portion, such as $\Delta$ %, of the coherence block. All frames are assumed to be received synchronously at the access points. This uncoordinated and simultaneous trans-mission of pilot sequences, without prior knowledge of active devices, constitutes the grant-free transmission mechanism. The system must therefore detect device activity and estimate channels jointly from the received signals at the access points. The received signal $\mathbf{y_{mn}} \in \mathbb{C}^{L\times 1}$ at the $n$-th antenna of the $m$-th access point can be expressed as :

$$\begin{aligned}\mathbf{y_{mn}} &= \sum_{k=1}^{K} a_k \rho_k^{1/2} g_{mk}^{(n)} s_k + w_{mn} \\ &= \mathbf{S D_a D_\rho}^{1/2} \mathbf{g_{mn}} + \mathbf{w_{mn}},\end{aligned} \quad (3)$$

where $\mathbf{S} = [\mathbf{s_1}\ \mathbf{s_2}\ \ldots\ \mathbf{s_K}] \in \mathbb{C}^{L\times K}$ represents the collection of all pilot sequences code book. The power transmitted by device $k$ is denoted as $\rho_k$, and the matrix $\mathbf{D_\rho} = \mathrm{diag}(\rho_1, \rho_2, \ldots, \rho_K)$ is the diagonal matrix containing these powers. The matrix $\mathbf{D_a} = \mathrm{diag}(\mathbf{a})$ represents the activity matrix. The channel vector $\mathbf{g_{mn}} = [g_{m1}^{(n)}, g_{m2}^{(n)}, \ldots, g_{mK}^{(n)}]^t \in \mathbb{C}^{K\times 1}$ describes the channel between all $K$ MTC devices and the $n$-th antenna of the $m$-th AP, where $[\cdot]^t$ indicates the transpose operation. Lastly, $\mathbf{w_{mn}} \sim \mathcal{CN}(0, \sigma^2 \mathbf{I}_L)$ is an $L$-dimensional complex gaussian random vector, representing noise at $n$-th antenna of the $m$-th AP. Thus, the signal $\mathbf{Y_m} \in \mathbb{C}^{L\times N}$ received at the $m$-th access point can be expressed as:

$$\mathbf{Y_m} = \mathbf{S D_a D_\rho}^{1/2} \mathbf{G_m} + \mathbf{W_m}, \quad (4)$$

where $\mathbf{G_m} = [\mathbf{g_{m1}}, \mathbf{g_{m2}}, \ldots, \mathbf{g_{mN}}] \in \mathbb{C}^{K\times N}$ is the channel matrix between all devices and the $m$-th AP, and $\mathbf{W_m} = [\mathbf{w_{m1}}, \mathbf{w_{m2}}, \ldots, \mathbf{w_{mN}}] \in \mathbb{C}^{L\times N}$ is an additive Gaussian noise matrix. Then, by omitting the index $m$, the received signal at any access point can be expressed as:

$$\mathbf{Y} = \mathbf{SX} + \mathbf{W}, \quad (5)$$

where $\mathbf{X} = \mathbf{D_a D_\rho}^{1/2} \mathbf{G_m} \in \mathbb{C}^{K\times N}$ is a row sparse matrix. The device activity detection problem aims to find the most likely activity vector given the code book of pilot sequences $\mathbf{S}$. This process involves identifying the indices of the nonzero entries (i.e., rows) in the sparse matrix $\mathbf{X}$. These indices represent the support of $\mathbf{X}$, *i.e.*, the locations of the non-zero rows. The support of matrix $\mathbf{X}$ can be defined as follows:

$$\mathrm{supp}(\mathbf{X}) = \left\{ k \in \mathbb{N}^+ \mid \mathbf{x}_{k,:} \neq 0 \right\}, \quad (6)$$

where $\mathbf{x}_{k,:}$ denotes the $k$-th row of the sparse matrix $\mathbf{X}$. we can denote that each column of row sparse matrix $\mathbf{X}$ shares the same locations of non-zero entries. Therefore, finding the support of $\mathbf{X}$ from the array received matrix $\mathbf{Y}$, comprising all received vectors, and the matrix $\mathbf{S}$ which is considered as the sensing matrix, can be seen as a sparse support recovery problem of the multiple measurement vector (MMV) model corrupted by additive noise. Consequently, the MMV problem can be formulated as follows [30]:

$$P_0 : \min_{\mathbf{X}} \|\mathbf{X}\|_0 \quad \text{s.t.} \quad \mathbf{Y} = \mathbf{SX}, \quad (7)$$

where, $\|\mathbf{X}\|_0 = |\mathrm{supp}(\mathbf{X})|$. The $P_0$-minimization problem is NP-hard due to the non-convexity of the L0-pseudo-norm. Common approximations, such as convex relaxation methods and iterative algorithms, as well as various other approaches, offer varying degrees of computational efficiency to tackle the $P_0$ computation complexity. The minimization problem $P_0$ can be addressed using DL, a specialized branch of machine learning (ML), particularly for detecting sporadic device activity in mMTC systems.

## IV. DEEP LEARNING FRAMEWORK

A general DL framework based on input I-Q received samples is proposed to address AD challenges in mMTC systems, with a particular focus on sparse recovery tasks. This framework typically employs NN architectures capable of capturing complex input-output relationships, with the possibility of utilizing multiple layers to enhance detection probability and efficiency. However, a specific architecture will be determined based on further analysis and experimentation.

Particularly, regression analysis using a DL architecture involves mapping an input signal to a specific output. A more capacity-efficient approach arises from recognizing that a sparse vector can be efficiently derived using the least squares estimator, notably when its support is known. Instead of directly predicting the activity vector $\mathbf{a}$, we propose using a DL model to approximate the mapping between $\mathbf{Y}$ and the support of row sparse matrix $\mathbf{X}$, which is the support of sparse activity vector $\mathbf{a}$. This intentional reformulation transforms the problem into a multi-binary label classification task, which is particularly effective for identifying active devices.

The following subsections detail the data generation process, the general adopted neural network architecture with its diverse input features for efficient AD, and the training phase.

### A. DATA GENERATION

The challenge of gathering large amounts of real data is a major problem when using DL in wireless communication systems. Fortunately, synthetic data can train DNN to tackle the $P_0$ minimization problem presented in (7). In this study, we prioritized accurately replicating real-world conditions during data generation phase. Initially, this required assigning MTC devices to a specified square area of $D \times D$ km$^2$, with APs placed to maintain realistic AP-AP and MTC device-AP distances, ensuring appropriate spacing, edge distance, and coverage distribution, as depicted in Fig. 3.

As the geographic topology was set up, the LSF effects were calculated using the 3GPP standard UMa criteria, incorporating the LoS condition. This was important because it highlighted how complex signal transmission is in metropolitan environments, where massive structures (distance attenuation, shadowing) have a significant impact on received signal behavior. Furthermore, the probability of each device being active was indicated by their activity status, calculated using a binomial distribution. As in [20], this study ensures that 95% of active devices gain access to the network by guaranteeing that all active devices achieve a target signal-to-noise ratio (SNR) at their dominant access point. The dominant AP can be defined as the access point that provides the strongest channel gain.

TABLE 3. Notations

| Notation | Description |
|---|---|
| $\tilde{a}$ | Predicted outputs vector of the DNN |
| $\hat{a}$ | Estimate of the activity descriptor in its binary domain |
| $a$ | Activity descriptor of the $K$ devices |
| $a_k$ | Activity status of the $k$-th device |
| $\tilde{a}_{Mk}$ | $M$ predicted activity values for device $k$ |
| $\tilde{a}_{Tk}$ | $T$ best predicted values for device $k$ |
| $\mathcal{A}$ | Set of active devices |
| $b_Z$ | Bias vector at layer $Z$ |
| $d'_{BP}$ | Break-point distance |
| $D_a$ | Devices activity matrix |
| $D_\rho$ | Devices powers matrix |
| $D_{\text{train}}$ | Training dataset |
| $D_{\text{test}}$ | Testing dataset |
| $F_{mk}$ | Shadow fading component between device $k$ and AP $m$ |
| $g_{mk}^{(n)}$ | Channel gain between device $k$ and antenna $n$ of AP $m$ |
| $G_m$ | Channel matrix between devices and AP $m$ |
| $h_{\text{BS}}$ | AP/base station height |
| $h_{\text{UT}}$ | Device height |
| $\mathbf{M}$ | Subset of all APs with $\|\mathbf{M}\| = M$ |
| $\mathbf{M}^\star_{\text{cen}}$ | Centralized trained model |
| $\mathbf{M}^\star_{\text{dec}}$ | Decentralized trained model |
| $PL$ | Path loss model |
| $s_k$ | Pilot sequence of device $k$ |
| $S$ | Pilot sequences codebook |
| $\mathcal{T}$ | Subset of APs with $\|\mathcal{T}\| = T$ |
| $U_z$ | Weight matrix between layer $z-1$ and $z$ |
| $w_{mn}$ | Noise vector at antenna $n$ of AP $m$ |
| $W_m$ | Gaussian noise matrix at AP $m$ |
| $y_{mn}$ | Signal at antenna $n$ of AP $m$ |
| $Y_i^{\mathcal{M}}$ | Signals from all APs at slot $i$ |
| $Y_i^{\mathcal{T}}$ | Signals from AP cluster $\mathcal{T}$ |
| $Y_{i,m}$ | Signal at AP $m$ at slot $i$ |
| $Y_m$ | Signal at AP $m$ |
| $x_{k,:}$ | Row $k$ of sparse matrix $X$ |
| $x^{(z)}$ | Output of layer $z$ |
| $X$ | Row-sparse matrix |
| $\alpha_{tk}$ | Weight for AP $t$ and device $k$ |
| $\rho_k$ | Transmit power of device $k$ |
| $\theta$ | DNN trainable parameters |
| $\beta_{mk}$ | LSF coefficient between AP $m$ and device $k$ |

To model the SSF, two approaches are available to generate rayleigh fading channels. The first approach allows fading to vary based on device activity, ensuring an accurate depiction of real conditions and evaluating the system's reaction to fluctuations. The second approach adopts a stable fading model for multiple device activities. Using previous approaches facilitates an in-depth evaluation of system performance under both dynamic and stable scenarios, yielding valuable insights into its overall performance.

Finally, the signal received at each AP was calculated using (4), which combined the transmitted pilot sequences, the large and small scale fading effects, and the additive Gaussian noise. Through this extensive data generation process, we developed a reliable dataset, which was key to evaluating the proposed algorithm's performance in realistic CF-mMIMO network simulations.

### B. DEEP NEURAL NETWORK

A DNN is a widely used form of DL that consists of artificial neural networks (ANN) with multiple layers between the input and output layers. The DNN architecture, shown in Fig. 4, includes an input layer that processes incoming data. This input data is defined as the signal representation, which can vary depending on the specific activity detection strategy being implemented. Table 4 provides a detailed outline of the different input signal representations used for both centralized and decentralized strategies within the AD framework.

The proposed architecture comprises $Z$ densely connected hidden layers, each containing $V$ nodes. These hidden layers are followed by a concatenation layer, which groups and resizes their outputs based on the input signal representation (as detailed in Table 4). The input to the architecture can be either a single signal $\mathbf{Y}$ or multiple signals $\mathbf{Y}_t$, where $\mathcal{T} \subset \{1, 2, \ldots, M\}$ denotes a subset of APs, and $t \in \mathcal{T}$ with $|\mathcal{T}| = T$. Here, $T$ represents the cluster size, which will be explained later. Each input type corresponds to a hidden layer output, which then feeds into the concatenation layer. Finally, the architecture includes an output layer consisting of $K$ neurons, responsible for predicting the activity descriptor vector.

Based on (16), the hidden layers in the model employ the rectified linear unit (ReLU) as their activation function because it is widely used due to its ability to introduce non-linearity while being computationally efficient, as it supports the use of back propagation for efficient learning in complex and deep network architectures.

$$ReLU(c) = \max(0, c). \tag{8}$$

In addition, the sigmoid activation function is employed in the output layer because it plays a crucial role in binary classification tasks. This decision is based on the particular ability to perform binary classification for each device in order to identify their activities, and it is provided by:

$$\text{Sigmoid}(c) = \frac{1}{1 + e^{-c}}. \tag{9}$$

The input and output dimensions of the DNN are primarily determined by several key criteria. The input depends on $L$, $N$, and $T$. On the other hand, the output is determined solely by $K$. Consequently, any changes in these parameters require retraining for the network to effectively adapt.

The weight matrices and bias vectors, which are the parameters of the DNN layers, are now introduced. We define the weight matrices according to the dimensions of the inputs and outputs of each layer as follows:

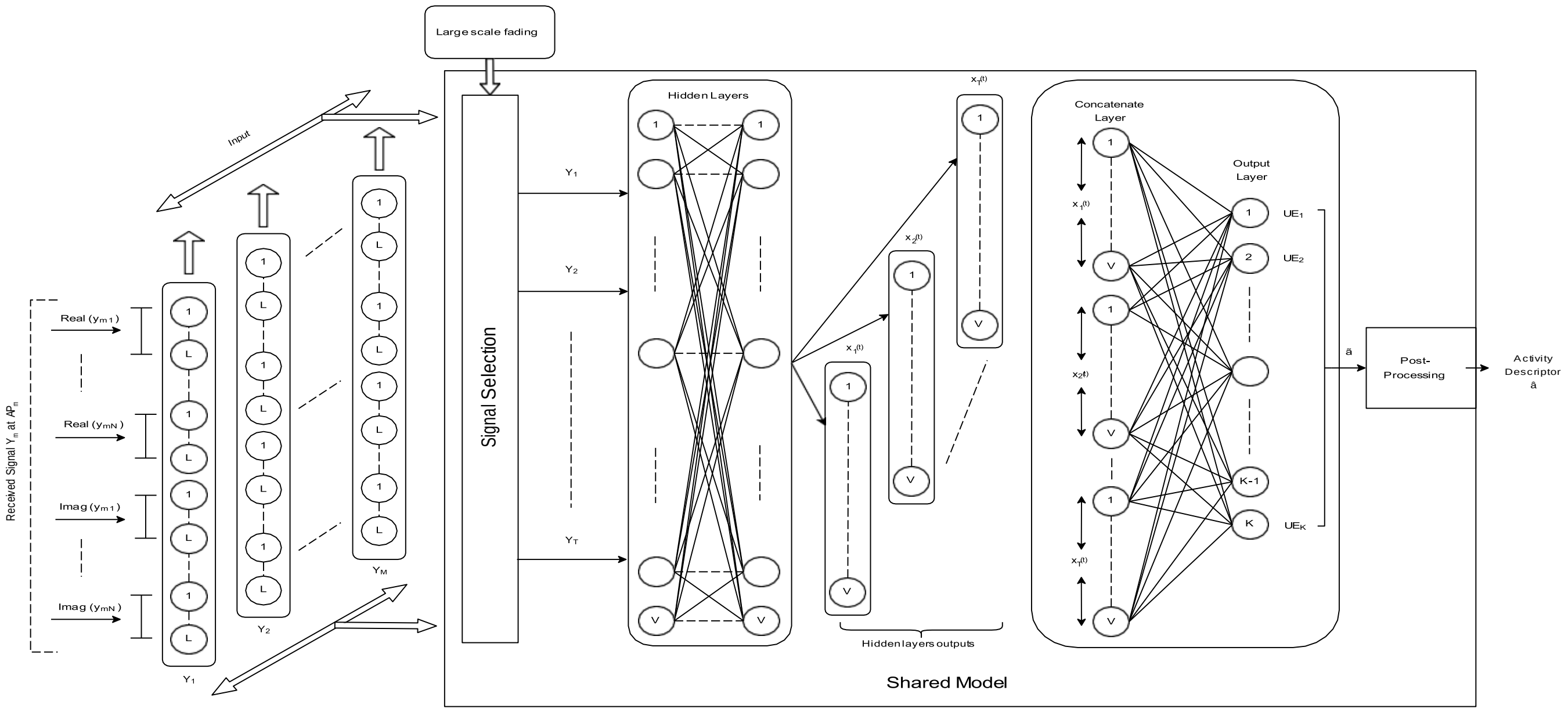

FIGURE 4. Architecture of the DNN algorithm proposed for device activity detection.

The weight matrix connecting the input layer to the first hidden layer is $\mathbf{U}_1 \in \mathbb{R}^{V\times 2NL}$, with the input determined by the strategy outlined in Table 4. Whether the input consists of a single signal or multiple signals, the same weight matrix is used. This ensures that the number of parameters remains constant, unaffected by the number of signals $T$ fed into the model.

For subsequent hidden layers, denoted as $1, 2, \ldots, Z$, the weight matrices are defined as $\mathbf{U}_2 \in \mathbb{R}^{V\times V}$, $\mathbf{U}_3 \in \mathbb{R}^{V\times V}, \ldots, \mathbf{U}_Z \in \mathbb{R}^{V\times V}$, where $Z$ represents the total number of hidden layers, and $\mathbf{U}_z$ denotes the weight matrix connecting layer $z-1$ to layer $z$. Following the hidden layers, a concatenation layer is introduced as a virtual layer that rearranges the outputs without adding any weights or trainable parameters. The output layer, denoted as $(Z+1)$, has a weight matrix $\mathbf{U}_{Z+1} \in \mathbb{R}^{K\times TV}$, where $T$ is the cluster size, and $V$ is the number of nodes per hidden layer. In the same way, the bias-valued vectors are $\mathbf{b}_1 \in \mathbb{R}^{V\times 1}$, $\mathbf{b}_2 \in \mathbb{R}^{V\times 1}, \ldots, \mathbf{b}_Z \in \mathbb{R}^{V\times 1}$ and $\mathbf{b}_{Z+1} \in \mathbb{R}^{K\times 1}$. Considering this, we use the DNN parameters to define the set by:

$$\theta = \{\mathbf{U}_z, \mathbf{b}_z\}, \quad z = 1, 2, 3, \ldots, Z \tag{10}$$

Finding the number of trainable parameters in the DNN method is feasible when one knows $\theta$. The number of trainable parameters is determined by taking into account the values of the bias vectors and the weights matrices:

$$\theta(K, L, N, V, Z, T) = (Z-1)V^2 + (2NL + TK + Z)V + K. \tag{11}$$

To process the outputs $\tilde{\mathbf{a}}$ of the DNN, which are probability values within the range $[0, 1]$, a post-processing block is introduced at the output of the algorithm. This block calculates the estimate of the activity descriptor in its original domain, denoted as $\hat{\mathbf{a}}$, as defined in Fig. 4.

In the process of detecting device activity, two primary types of errors can occur. The first is a false alarm (FA), which happens when a device that is not active is incorrectly identified as active. The second type is a miss detection (MD), occurring when an active device is wrongly classified as inactive. The likelihoods of both FA and MD are determined using a specific hard decision threshold $\tau$.

$$P_{\mathrm{FA}}(\tau) = \Pr(\hat{a}_k = 1 \mid a_k = 0) \tag{12}$$

$$P_{\mathrm{MD}}(\tau) = \Pr(\hat{a}_k = 0 \mid a_k = 1), \tag{13}$$

where $a_k$ denotes the true activity status of device $k$, and $\hat{a}_k$ represents the estimated activity. The probability of error within each random access slot is affected by the varying number of active devices, which in turn alters the frequency of specific error types. This variability is encapsulated in the error probability function that is expressed using the false alarm and miss detection probabilities related to the threshold $\tau$ and the probability of activation $\varepsilon$ as:

$$P_E(\tau, \varepsilon) = (1-\varepsilon)P_{\mathrm{FA}}(\tau) + \varepsilon P_{\mathrm{MD}}(\tau) \tag{14}$$

### C. TRAINING PHASE

In practical settings, our model operates initially in two distinct phases due to the absence of available training data. The initial phase, termed the transitional phase, entails the accumulation of the required data for model training, during which the model is trained on the collected dataset, following the stages outlined in algorithm 1, depending on the deployment strategy of the model (i.e., centralized or decentralized).

Subsequently, our research predominantly concentrates on the second phase, referred to as the permanent phase (i.e., model deployment for testing). This phase is detailed in section V. The simulations and results of this study are rooted in

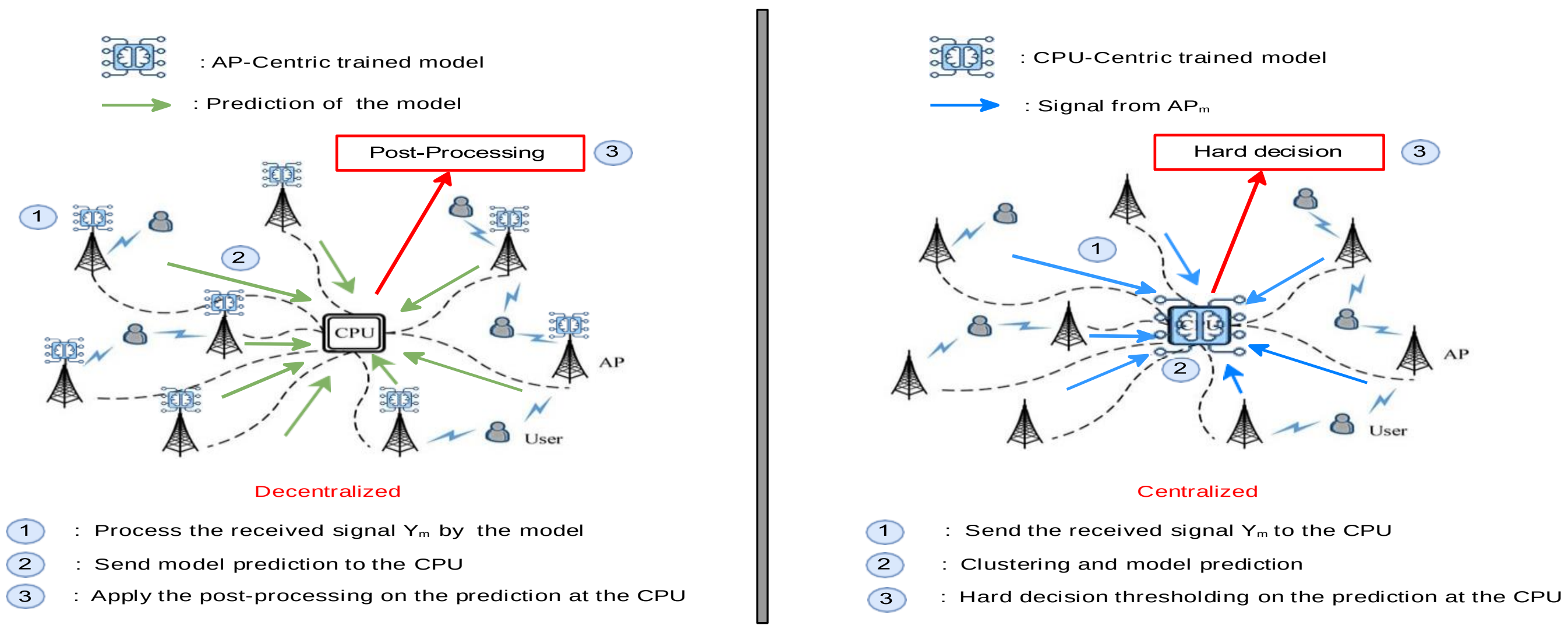

**FIGURE 5.** Signal exchange scheme.

---

**Algorithm 1** Activity detection strategies: Training procedure

---

**Input:** Training dataset $\mathcal{D}_{\text{train}} = \{(\mathbf{Y}_i^{\mathcal{M}}, \mathbf{a}_i)\}_{i=1}^{I}$
**Training Phase:**
**for** each random access slot $i = 1$ to $I$ **do**
  Let $\mathbf{Y}_i^{\mathcal{M}} = \{\mathbf{Y}_{i,1}, \ldots, \mathbf{Y}_{i,M}\}$
  **if** Centralized **then**
    - Randomly select a subset $\mathcal{T} \subset \{1, \ldots, M\}$ with $|\mathcal{T}| = T$
    - Form cluster input $\mathbf{Y}_i^{\mathcal{T}} = \{\mathbf{Y}_{i,m}\}_{m\in\mathcal{T}}$
    - Train model on $(\mathbf{Y}_i^{\mathcal{T}}, \mathbf{a}_i)$
  **else** {Decentralized}
    - Train model on full input one by one $(\mathbf{Y}_i^{\mathcal{M}}, \mathbf{a}_i)$
  **end if**
**end for**
**if** Centralized **then**
  - Store $\mathcal{M}^{\star}_{\text{cen}}$ on the CPU
**else** {Decentralized}
  - Store and deploy $\mathcal{M}^{\star}_{\text{dec}}$ over all APs
**end if**

---

the permanent phase, assuming that the model has progressed beyond the preliminary stage of data collection.

Let $\mathcal{D}_{\text{train}} = \{(\mathbf{Y}_i^{\mathcal{M}}, \mathbf{a}_i)\}_{i=1}^{I}$ be the training dataset synthetically generated with respect to the system model (see (4)), due to the lack of real data. Here $I = |\mathcal{D}_{\text{train}}|$ denotes its size. Each sample corresponds to a random access slot and consists of $\mathbf{Y}_i^{\mathcal{M}} = \{\mathbf{Y}_{i,1}, \ldots, \mathbf{Y}_{i,M}\}$, representing the received signals from all $M$ access points for a given device activity descriptor $\mathbf{a}_i$. As discussed, the absence of real data obliges us to resort to synthetic methods, as explained in Section IV-A. During the training phase, however, the data are generated from the system model and the knowledge of device activity is indispensable. In this phase, the activity descriptors (labels) are available, since supervised learning requires them. In contrast, during the inference phase, only the received signals are provided to the model, while the true labels are used solely for performance evaluation and metric computation.

Once the data are produced, we can train the model under two different strategies: centralized and decentralized. In both cases, the training occurs over multiple random access slots, denoted by $i = 1$ to $I$, where each slot contains the observations $\mathbf{Y}_i^{\mathcal{M}}$. The training phase is detailed in algorithm 1 as follows:

- **Centralized strategy:** At each slot, a subset of APs $\mathcal{T}$ of fixed size $T$ is randomly selected from the full observations $\mathbf{Y}_i^{\mathcal{M}}$. The training input is then formed from these observations as $\mathbf{Y}_i^{\mathcal{T}}$, representing the selected APs. This process is repeated over all access slots, and the model is trained on these inputs with the corresponding activity label $\mathbf{a}_i$, resulting in a centralized trained model denoted as $\mathcal{M}^{\star}_{\text{cen}}$.
- **Decentralized strategy:** The training is performed directly on the complete set of observations $\mathbf{Y}_i^{\mathcal{M}}$, one by one with the corresponding activity label $\mathbf{a}_i$, without any selection. After training is completed across all access slots, the resulting trained model for the decentralized deployment is denoted by $\mathcal{M}^{\star}_{\text{dec}}$.

The optimization problem can be stated as follows:

$$g(Y, S, \theta) = \min_{\mathbf{X}} \|\mathbf{Y} - \mathbf{S}\mathbf{X}\|_F, \tag{15}$$

where $\|\cdot\|_F$ is the frobenius norm, and the model parameters are indicated by $\theta$. A DNN is trained to map each input $Y_i$ to a desired estimated activity indicator $\mathbf{a}_i$ via several successive layers of linear transformation interspersed with element-wise nonlinear transforms, given a collection of training examples $D_{train}$. The input-output relationship for the hidden layers in a standard feed-forward neural network (FFNN) is expressed as follows:

$$\mathbf{x}^{(z)} = f\left(\mathbf{U}_z \mathbf{x}^{(z-1)} + \mathbf{b}_z\right), \tag{16}$$

where $f(\cdot)$ denotes the non-linear activation function, as defined in (8) and (9). In this equation, $\mathbf{x}^{(z)}$ represents the output of the layer $z$, with $\mathbf{x}^{(z-1)}$ being the input received from the previous layer $z-1$. The parameters of the network at this stage, namely the weight matrix $\mathbf{U}_z$ between layer $z-1$ and layer $z$, and the bias vector $\mathbf{b}_z$ at layer $z$, are collectively represented by $\theta$.

These parameters are learned and updated during training utilizing the adam optimizer. This optimizer, a popular variant of stochastic gradient descent, dynamically adjusts the learning rate $\eta$ for each parameter. The adam optimizer is particularly popular in ML due to its adaptability to handle sparse gradients and use of an adaptive learning rate mechanism, which allows it to converge in complicated scenarios with resilience. We denote the output of the DL algorithm for the input sample $(\mathbf{Y}^{\mathrm{M}}_i, \mathbf{a}_i)$ as $\tilde{\mathbf{a}}_i$. For the network training, we use the binary cross-entropy loss [31] calculated over all the $I = |D_{train}|$ samples:

$$\mathcal{L}(D_{train}) = -\sum_{i=1}^{I} \left[\mathbf{a}_i \log(\tilde{\mathbf{a}}_i) + (1-\mathbf{a}_i)\log(1-\tilde{\mathbf{a}}_i)\right]. \tag{17}$$

Using the back-propagation technique, we may propagate the loss all the way and train the parameters of the network. The loss function $\mathcal{L}(D_{train})$ is minimized during the training process by choosing the right parameters $\theta$. If after multiple epochs the loss function in (17) does not improve, the training ends. That is, the trained DL network estimates the activity descriptor with high detection probability and a low false alarm rate.

There is an offline option for the training procedure. After the DNN parameters are learned, it is computationally less expensive to use this neural network for identifying active devices in new datasets (inference phase). The size of the new dataset for inference is typically 20% of the training data size, denoted as $D_{\text{test}}$ with size $J = |D_{\text{test}}|$. During the inference phase, the neural network only requires a few vector-matrix multiplications, summations, and element-wise nonlinear operations.

It is important to note that the proposed model is conditioned on several system-level parameters, such as the number of antenna elements, pilot sequence length, cluster size, and device density. Since these parameters are embedded in the model architecture and training process, any variation in their values may necessitate fine-tuning or, in some cases, retraining of the model. As an example, when new devices join the network, new pilot sequences must be assigned to accommodate them, which can lead to a change in the number of active devices and consequently alter the size of the model output layer making retraining necessary to adapt the model to the new configuration.

**Algorithm 2** Activity detection strategies: Inference procedure

**Setup:** Load trained models $\mathcal{M}^\star_{\text{cen}}$ or $\mathcal{M}^\star_{\text{dec}}$
**Loop: For each access slot $j$ (real-time)**
**if** Centralized **then**
  **for** each AP $m = 1$ to $M$ **do**
    - Receive signal $\mathbf{Y}_{j,m}$ from AP $m$
    - Forward $\mathbf{Y}_{j,m}$ to the CPU
  **end for**
  At CPU:
  **for** each device $k = 1$ to $K$ **do**
    - Select signals from the dominant AP cluster based on LSF for device $k$, denoted as $\mathbf{Y}^{\mathrm{T}}_j$
    - Predict activity: $\tilde{a}_k = \mathcal{M}^\star_{\text{cen}}(\mathbf{Y}^{\mathrm{T}}_j)$
  **end for**
  - Apply hard decision mechanism on the predicted activity vector $\tilde{a}$ to obtain final activity vector $\hat{a}$
**else** {Decentralized}
  **for** each AP $m = 1$ to $M$ **do**
    - Receive signal $\mathbf{Y}_{j,m}$
    - Predict local descriptor $\tilde{\mathbf{a}}_m = \mathcal{M}^\star_{\text{dec}}(\mathbf{Y}_{j,m})$
    - Send $\tilde{\mathbf{a}}_m$ to CPU
  **end for**
  At CPU:
    - Cluster APs per device based on LSF
    - Aggregate predictions from relevant APs cluster
    - Apply post-processing and output $\hat{a}$
**end if**

## V. MODEL DEPLOYMENT AND INFERENCE STRATEGIES

This phase addresses the inference process of the model under both deployment strategies: centralized and decentralized.

After the completion of the training phase, the trained model is deployed and evaluated using the testing dataset $D_{\text{test}} = \{(\mathbf{Y}^{\mathrm{M}}_j, \mathbf{a}_j)\}_{j=1}^{J}$, whose size corresponds to 20% of the training dataset, where $j$ represents the access slot index.

Although the inference process is intended to operate in real deployment scenarios, as denoted in the algorithm 2 where real-time detection at APs is performed, in this study we apply it on the testing dataset to facilitate performance evaluation and benchmarking.

The section also presents the clustering approach employed during inference, along with a detailed description of both deployment strategies.

### A. CLUSTERING-BASED ACTIVITY DETECTION

Traditional activity detection algorithms rely on data from a single AP per device. In CF-mMIMO networks, utilizing all APs for activity detection is more effective but computationally expensive.

A clustering-based approach for GF-RA, as proposed in [20], groups APs with strong channels to each device, requiring accurate estimates of the LSF coefficients $\beta_{mk}$. These coefficients are used to determine the best cluster for each device and within the model to predict activity by solving a polynomial equation of degree $2T - 1$, where $T$ is the number of APs per device cluster. While efficient, this method becomes computationally intensive as the number of devices and cluster sizes increase.

**TABLE 4.** Comparison of centralized and decentralized activity detection strategies

| | Activity Detection Framework | |
|---|---|---|
| | **Strategy I** | **Strategy II** |
| **Detection framework** | Decentralized | Centralized |
| **Model deployment type** | AP-centric | CPU-centric |
| **Signal selection** | Disabled | Enabled |
| **Input signal representation** | $Y_m$ | $Y_t$ / $t \in \mathcal{T} \subset \{1, 2, \ldots, M\}$, $\lvert\mathcal{T}\rvert = T$ |
| **Signal exchange scheme** | Prediction only | All AP-signals |
| **Clustering approach** | Post-model prediction | Pre-model prediction |
| **Post-processing method** | Processing I | Hard decision mechanism |
| | Processing II | |

The methodology presented refines the previously stated approach by maintaining the optimal cluster selection per device while decoupling the clustering framework from the DNN model learning phase. The dynamic cluster selection is performed by the CPU in both strategies, independently of the model, to enhance detection probability. Specifically, clustering is applied post-prediction in the first strategy and pre-prediction in the second. This separation improves flexibility and efficiency in activity detection for CF networks, significantly reducing computational complexity.

### B. STRATEGY I: DECENTRALIZED

In this approach, the steps of the model are outlined in Table 4, detailed in algorithm 2, and illustrated on the left side of Fig. 5. The framework operates as follows:

- The model functions in a distributed manner, with a copy of the trained model $\mathcal{M}^{\star}_{\text{dec}}$ deployed on each of the $M$ APs in the network.
- The architecture in Fig. 4 is simplified by disabling the signal selection block. Each AP $m$ directly uses the signal it receives as input.
- At AP $m$, the model processes the received signal $\mathbf{Y}_m = [\mathbf{y}_{m1}, \mathbf{y}_{m2}, \ldots, \mathbf{y}_{mN}] \in \mathbb{C}^{L\times N}$, consisting of the real and imaginary components of signals at each antenna $n$. This input is passed through the hidden layers, producing a single extracted feature. Since there are no multiple inputs, the concatenation layer has no effect, and the output layer predicts the activity descriptor vector $\tilde{\mathbf{a}}$ for all devices.
- This process is independently performed by all APs, resulting in $M$ activity descriptor vectors.
- No clustering is applied at this stage. As shown in Fig. 5, the $M$ predicted vectors are exchanged with the CPU. At the CPU, each device $k$ will have $M$ predicted activity values, represented as $\tilde{\mathbf{a}}_{\mathbf{Mk}} = (\tilde{a}_{1k}, \tilde{a}_{2k}, \ldots, \tilde{a}_{Mk})$.
- A post-prediction clustering process is then applied. For each device $k$, the CPU selects the $T$ best predicted values $\tilde{\mathbf{a}}_{\mathbf{Tk}} = (\tilde{a}_{1k}, \tilde{a}_{2k}, \ldots, \tilde{a}_{Tk})$ out of $M$, based on the large-scale fading coefficients $\beta_{mk}$ known at the CPU. These predicted values fall within the range $[0, 1]$.
- Using the $T$ selected predicted values, a post-processing method determines the final estimated activity of device $k$, denoted as $\hat{a}_k$.

Two different methods are employed in this approach for post-processing:

**Processing I:** This method consists of two blocks a hard decision phase followed by a fusion rule function.

- The hard decision mechanism applies a threshold $\tau \geq 0$ to binarize the $T$ predicted values for device $k$:

$$\bar{a}_{tk} = \begin{cases} 0, & \text{if } \tilde{a}_{tk} < \tau, \\ 1, & \text{if } \tilde{a}_{tk} \geq \tau. \end{cases} \tag{18}$$

  The binary outputs for device $k$ from its best cluster are represented as $\bar{\mathbf{a}}_{Tk} = (\bar{a}_{1k}, \bar{a}_{2k}, \ldots, \bar{a}_{Tk})$.
- The binary outputs are then processed by a fusion rule function, which can be AND, OR, or Majority. Majority rule is chosen for its balanced performance, where a device is considered active if the majority ($T/2$) of APs in the cluster detect activity. The output of the fusion rule provides the final estimated activity $\hat{a}_k$ of device $k$.

**Processing II:** This method consists of two blocks a ponderation function block followed by a hard decision phase.

- The goal of the ponderation function is to calculate a final prediction $\dot{a}_k$ for device $k$, which combines multiple predictions made by a cluster of $T$ APs. These predictions are weighted based on the quality of the channel between the device and each AP. The final prediction is given by:

$$\dot{a}_k = \sum_{t=1}^{T} \alpha_{tk} \tilde{a}_{tk}, \tag{19}$$

  where $\alpha_{tk}$ represents the weight assigned to the $t$-th AP for device $k$, and $\tilde{a}_{tk} \in [0, 1]$ denotes the predicted activity of device $k$ from the $t$-th AP in the cluster, as obtained by the decentralized model $M^{\star}_{dec}$, where $t = 1, 2, \ldots, T$.

- The weight $\alpha_{tk}$ reflects the LSF quality (signal strength or reliability) between the $t$-th AP and device $k$. To ensure the contributions of all APs in the cluster sum to 1, the weight is normalized as follows:

$$\alpha_{tk} = \frac{\beta_{\text{best},tk}}{\sum_{t=1}^{T} \beta_{\text{best},tk}}. \tag{20}$$

- The hard decision mechanism applies a threshold $\tau \geq 0$ to binarize the weighted outputs for each device $k$, denoted as $\dot{a}_k$ to provide the final estimated activity $\hat{a}_k$ of device $k$:

$$\hat{a}_k = \begin{cases} 0, & \text{if } \dot{a}_k < \tau, \\ 1, & \text{if } \dot{a}_k \geq \tau. \end{cases} \tag{21}$$

### C. STRATEGY II: CENTRALIZED

In this approach, the steps of the model are outlined in Table 4, detailed in algorithm 2, and shown on the right side of Fig. 5. The framework functions as follows:

- The model is centralized and runs on the CPU, independent of the APs in the network.
- The model architecture, shown in Fig. 4, includes the signal selection block.
- The exchange scheme requires APs to relay signals to the CPU, with input comprising signals from all $M$ APs in the network, as shown on the right side of Fig. 5.
- The centralized trained model $M^{\star}_{cen}$ processes signals in parallel for each device, utilizing the selection block to leverage large-scale fading coefficients to identify the best cluster of AP signals for each device. Specifically, it selects the top $T$ signals for device $k$, represented as $[\mathbf{Y}_1, \mathbf{Y}_2, \ldots, \mathbf{Y}_T]$, as illustrated in Fig. 4.
- The model processes the signals through the hidden layers, producing $T$ extracted feature vectors. These are concatenated in the concatenation layer to detect the activity for all $K$ devices, as each device has its own cluster of AP signals (though some clusters may overlap with different devices activities). It is essential to detect activity for all $K$ devices within their respective clusters. For device $k$, the model selects their specific activity by ignoring the activities of the remaining $K - 1$ devices, resulting in the predicted activity $\tilde{a}_k$, and ultimately producing the overall activity vector $\tilde{\mathbf{a}}$.
- Finally, the CPU applies a hard decision mechanism with a threshold $\tau \geq 0$ to binarize the predicted values, yielding the final estimated activity vector $\hat{\mathbf{a}}$.

## VI. NUMERICAL RESULTS

This section explores the topic of improved connectivity in CF-mMIMO architectures, focusing on the evaluation and comparison of various activity detection algorithms in both cell-free and cellular mMIMO systems under near-realistic scenarios. The investigation begins with mathematically driven algorithms, including AMP, ISTA, and FISTA, as detailed in [25]. It then transitions to a data-driven approach leveraging neural networks, specifically the DNN architecture, employed thus far. To comprehensively evaluate the effectiveness of activity detection, we utilize the widely recognized receiver operating characteristics (ROC) curve, which represents the probability of detection ($P_D = 1-P_{MD}$) against the probability of false alarm ($P_{FA}$). This method provides a robust framework for assessing detection performance across different operational thresholds.

TABLE 5. Simulation parameters.

| Parameter | | CF | Cellular |
|---|---|---|---|
| **System** | | | |
| Area size ($D^2$) | | 1 $Km^2$ | |
| Edge distance | | 50 m | |
| Number of devices (K) | | 100 | |
| Number of APs (M) | | 20 | 1 |
| Number of antennas (N) | | 2 | 8 |
| Pilot length (L) | | 40 | |
| Minimum distance: Device-AP | | 10 m | |
| AP Spacing | | 15 m | |
| Antenna height | | 12 m | |
| Device height | | 1.5 m | |
| Carrier frequency (fc) | | 900 MHz | |
| Cluster size (T) | | 4 | - |
| Sparsity level ($\varepsilon$) | | 0.1 | |
| Target SNR | | 42 dB | |
| **Coherence Block** | | | |
| Coherence time ($\tau_c$) | | 1 ms | |
| Coherence bandwidth ($B_c$) | | 200 kHz | |
| Number of symbols ($T_c$) | | 200 | |
| Reserved portion ($\Delta$) | | 20% | |
| **Channel Fading** | | | |
| Noise power ($\sigma^2$) | | -109 dBm | |
| Small scale fading | | Rayleigh | |
| Large scale fading | Path loss | 3GPP standard | |
| | Shadowing | $N(0, 1)$ | |
| **Neural Network Setup** | | | |
| Training data size ($I$) | | $5 \times 10^4$ samples | |
| Testing data size | | $10^4$ (20% of $I$) | |
| Batch size | | 256 | |
| Number of epochs | | 200 | |
| learning rate ($\eta$) | | 0.01 | |

### A. SIMULATION MODEL

Our study covers a detailed scenario for both cell-free and cellular networks, defined by a diverse set of parameters. The key parameters for these scenarios are given in Table 5. [2]

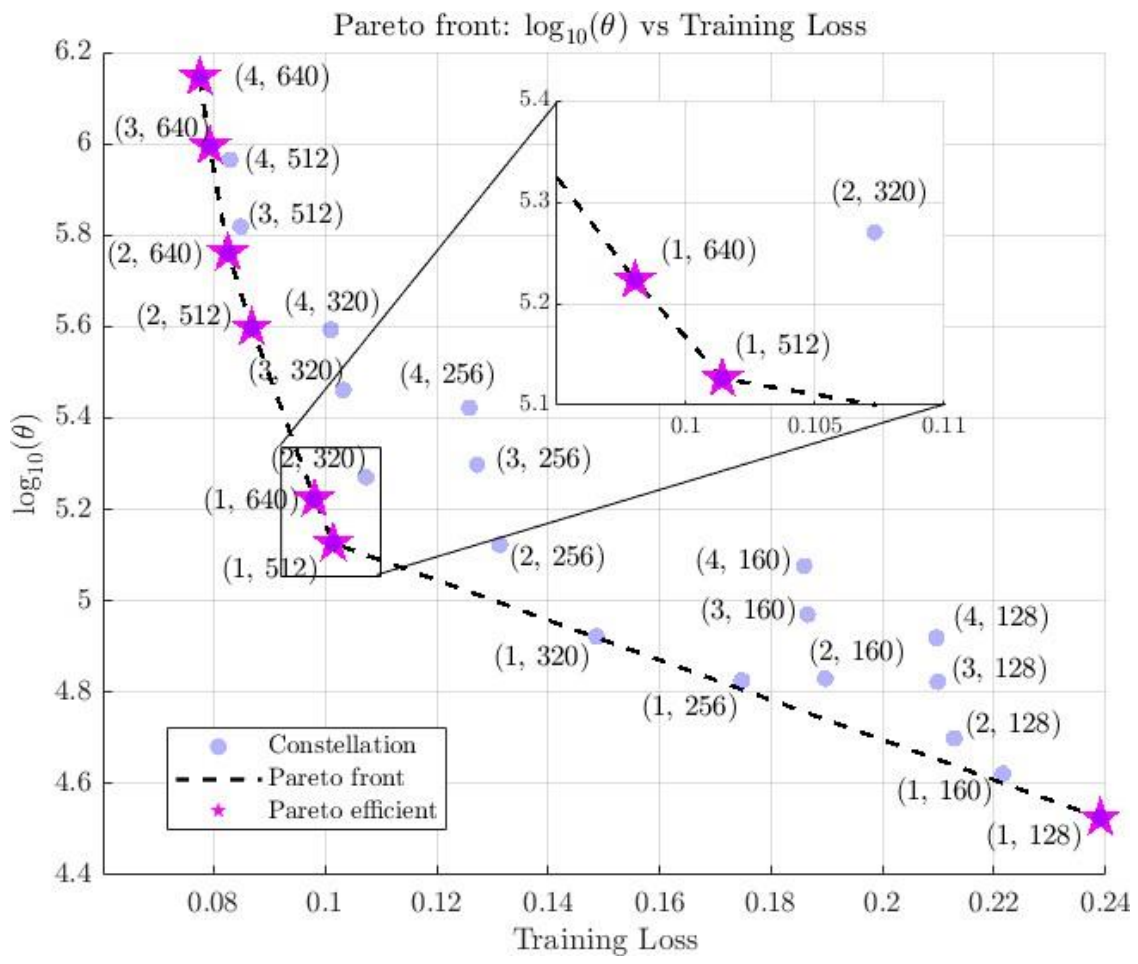

**FIGURE 6.** Pareto front : number of trainable parameters in the DNN vs training loss.

### B. DNN PARAMETERS TUNING: PERFORMANCE VS. COMPLEXITY

In this section, our primary focus is to evaluate the performance of DNN in the context of device activity detection in CF-mMIMO systems.

To achieve this goal and for simplicity, we conducted a thorough investigation into the architectural parameters of DNN, focusing initially on the decentralized approach. Our objective is to achieve a high detection probability in activity detection while mitigating the risk of overfitting.

The tuning process involves careful consideration of the number of neurons within the hidden layers of the DNN architecture. We have conducted a thorough parameter search, varying the number of neurons $V$ within the range {128, 160, 256, 320, 512, 640} and exploring different detector configurations with various numbers of intermediate hidden layers $Z$ within the set {1, 2, 3, 4}. This meticulous parameter-tuning process is an essential step to optimize the DNN for accurate activity detection.

To objectively study the performance-complexity trade-off, the detectors are compared based on pareto efficiency. A detector $(Z_i, V_j)$ is considered pareto-efficient if no alternative detector exists that reduces either the training losses, measured using binary cross-entropy on the training dataset in (17), or the complexity, assessed as the number of training parameters in (11), without adversely affecting the other metric. Since we are in the decentralized approach and, as previously stated, the concatenation layer has no effect, this scenario corresponds to the case where $T = 1$ in the equation (11). The set of all pareto-efficient detectors, known as the pareto front, represents the available trade-off options. Switching from one pareto-efficient detector to another involves prioritizing either complexity or training loss. Conversely, a detector that is not pareto efficient should not be selected, as it is possible to improve at least one of the metrics without compromising the other.

Analyzing the curve in Fig. 6, it is evident that increasing the number of parameters consistently decreases the minimum training loss. This trend is depicted in Fig. 6, where it is noticeable that increasing, the number of layers results in only a slight improvement in the validation loss for each value $V$. However, a substantial decrease in loss is observed when the number of neurons is increased for each $Z$ value.

Fig. 6 illustrates that the optimal trade-off between complexity and loss is achieved with the detector configured as $V = 512$ and $Z = 1$, as indicated by pareto-efficiency. Consequently, the general DNN architecture is simplified to a lightweight SLP, which will be utilized for the remainder of the simulations in both the centralized and decentralized approaches.

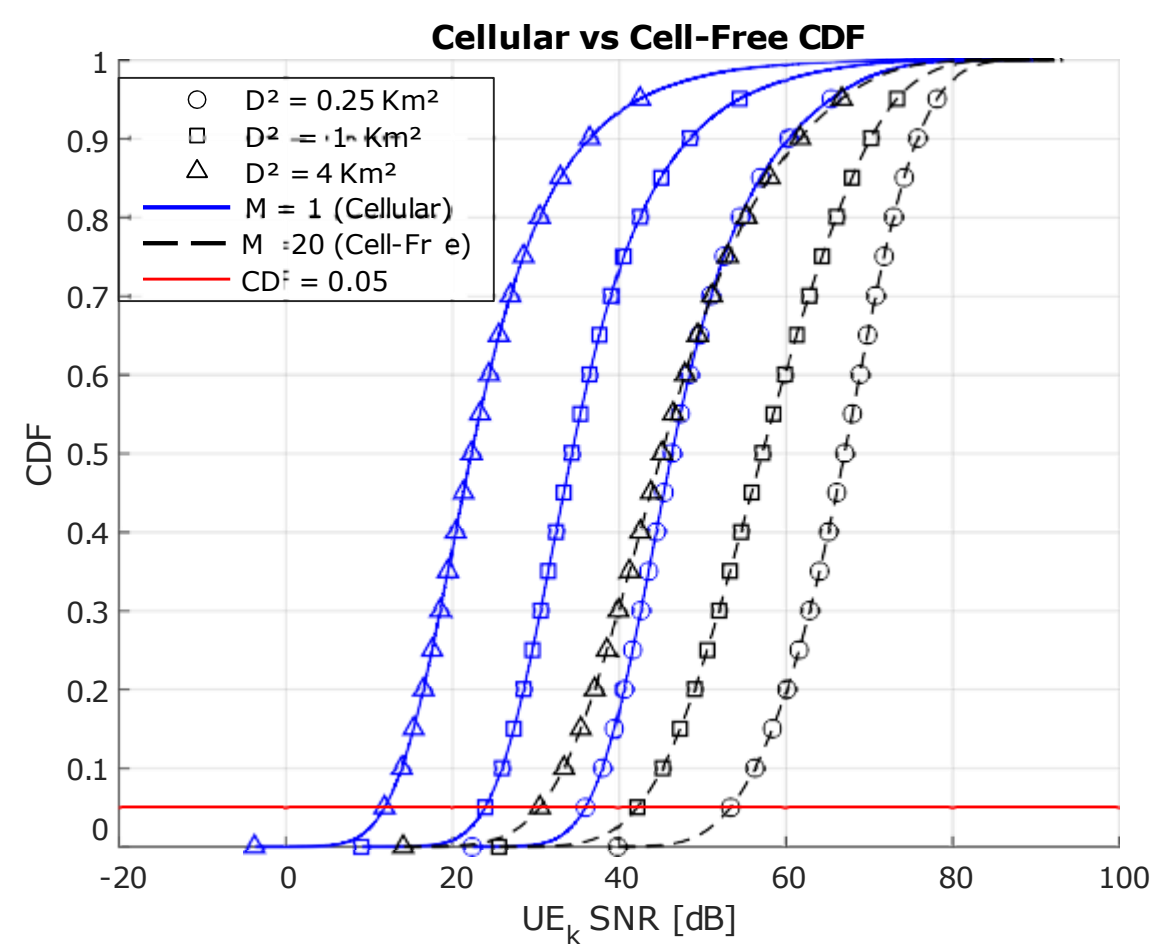

**FIGURE 7.** Active device SNR.

### C. SNR ANALYSIS RESULTS

Fig. 7 presents the cumulative distribution function (CDF) of the SNR for an active device in both a CF-mMIMO and a cellular network, based on the respective scenarios outlined in Table 5. The SNR is evaluated for devices transmitting at 200 mW in three coverage area sizes: 0.25 km$^2$, 1 km$^2$, and 4 km$^2$, and in two configurations: 1 AP (cellular) and 20 AP (cell-free). For the cell-free case, the SNR corresponds to the dominant AP.

It is evident from Fig. 7 that the SNR achieved in the CF network, represented by dashed lines, significantly outper-forms that of the cellular network, depicted by solid lines. This improvement indicates a lower likelihood of outage probability, translating to more reliable service for a larger number of devices. The enhanced SNR can be attributed to the carefully designed topology of access points and the realistic environmental model, which ensures devices are typically closer to an access point, benefiting from stronger channel gains and improved signal quality.

[2] See Section III for a comprehensive analysis of simulation assumptions and parameters. Briefly, with $\tau_C = 1$ ms and $B_C = 200$ kHz yielding $T_C = 200$ symbols and given $K = 100$ devices, orthogonal pilot sequences might be deemed feasible. However, in our simulations, we employ non-orthogonal pilot sequences, even though $K$ does not exceed $T_C$ to improve the timing efficiency of our simulations. Given our objective to expand connectivity, this strategy allows for a potential increase in the number of devices. This approach is justified by its minimal effect on the outcomes.

For simulation purposes, the SNR target at the dominant AP was set to ensure that 95% of the active devices meet the required SNR threshold to gain network access. This SNR target is depicted by the intersection of the plot lines with the reference line marked $CDF = 0.05$.

This analysis highlights the benefits of distributed AP deployment in the cell-free scenario, showcasing its ability to improve connectivity and minimize the risk of signal outages, which are essential for achieving robust and reliable performance. The use of markers distinguishes network configurations based on area size, underscoring the influence of this variable on SNR and overall network efficiency.

### D. ALGORITHM PERFORMANCE WITH VARIABLE SYSTEM PARAMETERS

In this section, we evaluate the performance of our proposed algorithms in both centralized and decentralized approaches using various post-processing methods. The evaluation includes a comparison of all algorithms, including our models and those from the literature [14], [25], across different system parameters. Specifically, we analyze the pilot sequence length $L$, the total number of devices $K$, and the sparsity parameter $\epsilon$. All models are assessed under the scenarios outlined in Table 5. These parameters are critical for optimizing the system and provide valuable insights into the algorithms' adaptability and scalability in real-world deployments.

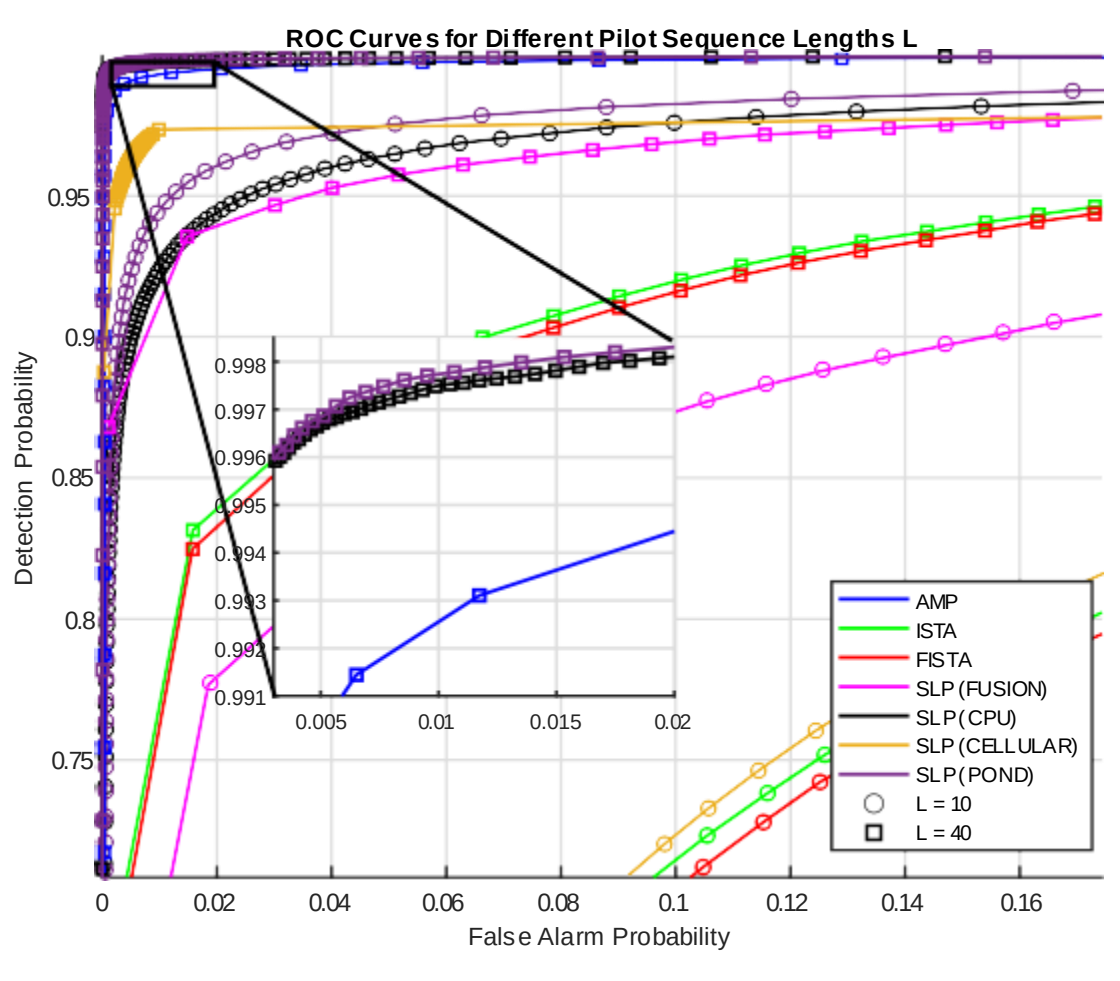

FIGURE 8. ROC curve analysis for different pilot sequence lengths $L$.

Fig. 8 presents the ROC curves for varying the pilot sequence lengths $L$, with all other parameters fixed as specified in Table 5. The results show a clear improvement in performance as the pilot length increases from 10 to 40. This enhancement is expected, as longer pilot sequences reduce the correlation between non-orthogonal pilots assigned to a fixed number of devices, improving device discrimination under the SLP algorithm. The plots also demonstrate that device activity detection performance is superior in cell-free massive MIMO networks compared to co-located MIMO networks, particularly as the pilot length increases. While the AMP algorithm struggles with shorter pilot lengths, it catches up as it $L$ grows, though at a higher computational cost. Moreover, the SLP algorithm, in both the centralized CPU and decentralized cases using weighting (POND), outperforms state-of-the-art algorithms across all values of $L$. For instance, with $L = 40$, the SLP achieves a detection probability $\geq 0.996$ with a false alarm rate of $0.005$, whereas the AMP algorithm yields a lower detection probability $< 0.991$ under the same false alarm constraint. However, longer pilot sequences occupy more of the coherence block, improving detection by lowering sequence correlation but reducing the symbols available for uplink and downlink data transmission. This trade-off may impact the overall data rate and increase the cost of transmission bandwidth.

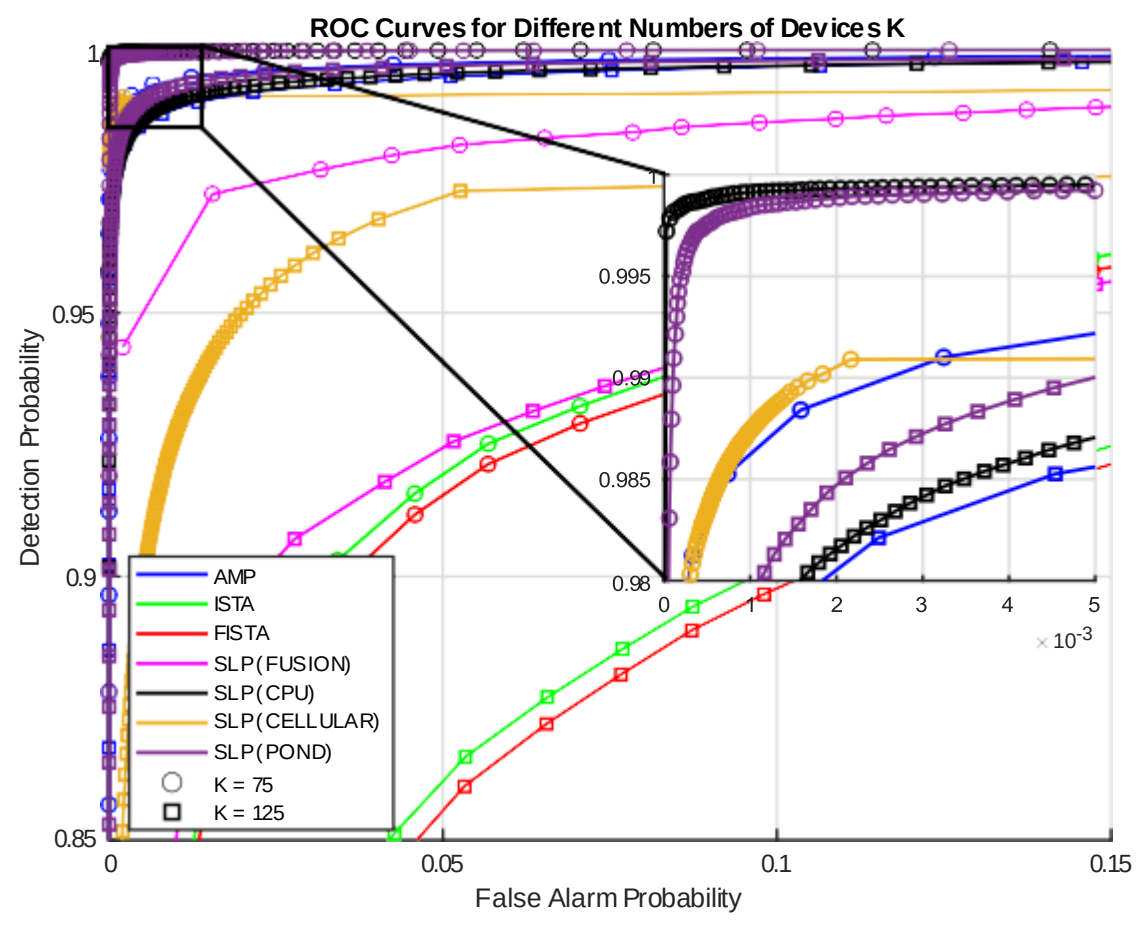

FIGURE 9. ROC curve analysis for different device numbers $K$.

Fig. 9 examines the performance of the models as the number of devices $K$ increases, while keeping all other system parameters fixed. As shown, performance tends to degrade with higher device density. This behavior is expected, as increasing $K$ enhances the correlation between pilot sequences due to more devices sharing a fixed resource pool, with the pilot sequence length held constant. In the model architecture (Fig. 4), the number of devices corresponds to the number of neurons in the last layer. A static input size combined with a high device count makes it more challenging for the network to differentiate between device signals, reducing discrimination performance. The results also highlight that CF networks outperform co-located networks across all scenarios. Both the centralized and decentralized (POND) approaches surpass state-of-the-art algorithms as the network becomes denser, demonstrating better scalability and robustness in high-density deployment scenarios. A general comparison of different post-processing methods in the decentralized approach reveals that majority fusion decision yields lower performance compared to weighted methods. Assigning weights to access points based on priority (POND) or treating all APs equally, as in the fusion case (FUSION), significantly affects the performance.

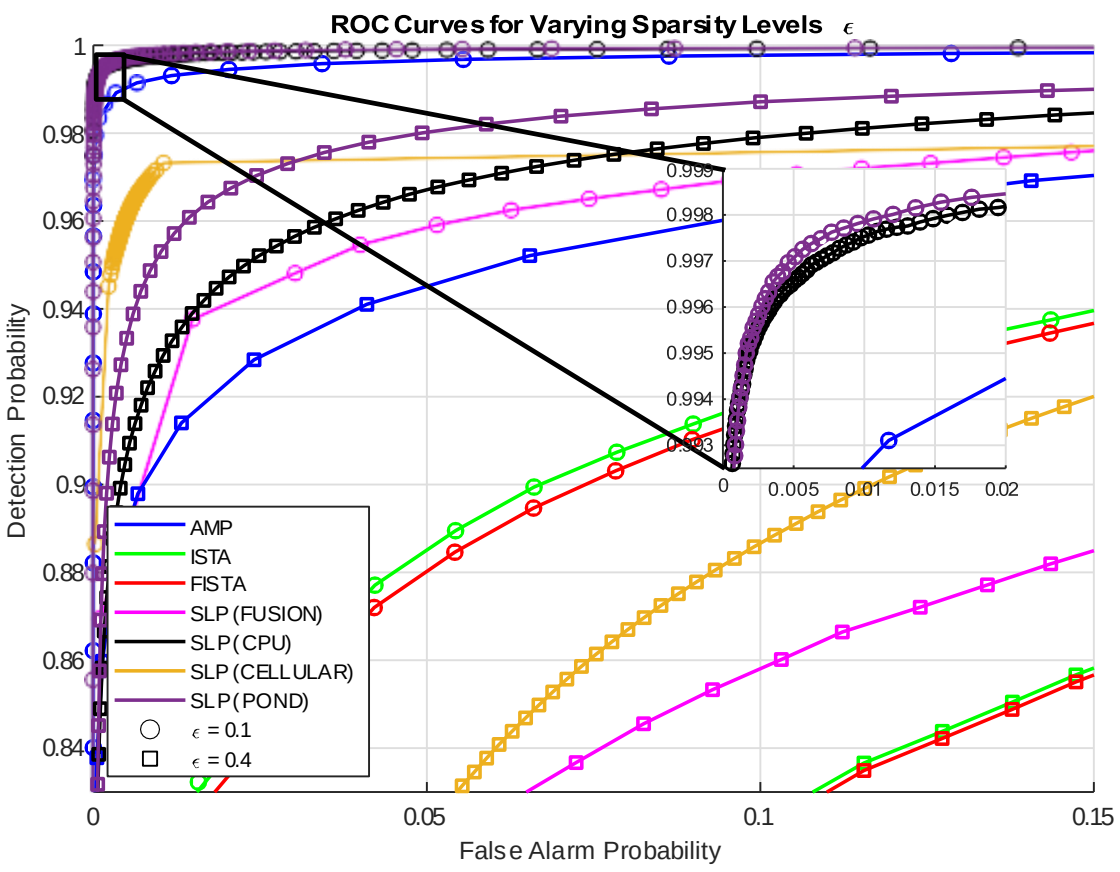

**FIGURE 10.** ROC curves for different sparsity parameters $\epsilon$.

Above all, these scenarios illustrate the under-sampling ratio ($\frac{L}{K}$), or the trade-off between the number of devices $K$ and the length of the pilot sequence $L$. The performance of the algorithm can be optimized by establishing a balance between these variables, highlighting the significance of a comprehensive system design that takes into account the device density and available sequence lengths for a best system operation.

These two scenarios highlight how system performance evolves with varying device density and pilot sequence length. While the proposed model demonstrates robust performance under such variations, its scalability in ultra-dense and mobile environments may still be challenged by factors such as increased pilot collisions and correlation, fixed pilot sequence lengths, and dynamic AP-user associations (e.g., clustering). Notably, the current design assumes quasi-static device positions, which may limit its effectiveness in high-mobility scenarios, where rapid fluctuations in large-scale fading can degrade inference performance. Addressing these challenges may require mobility-aware adaptation mechanisms. Furthermore, future extensions could explore adaptive pilot allocation strategies to mitigate scalability issues in ultra-dense deployments.

In Fig. 10, we analyze the performance of the models across varying sparsity levels $\epsilon$. As shown, performance decreases as $\epsilon$ increases, which is expected due to higher activity levels leading to greater interference. The CF scenario consistently outperforms the cellular counterpart, demonstrating its superior capability in managing sparse activity detection. The SLP with ponderation post-processing function achieves the best performance across all sparsity levels, surpassing all other algorithms. At lower $\epsilon$ values, both centralized and decentralized approaches are highly effective detectors. At higher values of $\epsilon$, the number of active devices increases, making the state more complex due to the appearance of more pilot sequences in the received signal. In this case, the sparsity level of the activity vector is reduced. Greedy pursuit algorithms such as AMP, ISTA, and FISTA are commonly used for sparse recovery; however, their performance degrades as the sparsity level decreases. In contrast, learning-based methods demonstrate superior robustness under such high-activity regimes.

Finally, the results emphasize the significance of employing both centralized and decentralized approaches with different post-processing methods. In the decentralized approach, the model learns features from a single received signal, whereas the centralized method processes a cluster of signals collected from the best APs. The decentralized approach with weighting demonstrates slightly better performance than the centralized CPU approach by leveraging external diversity through signal weighting, while the CPU approach explores diversity internally within the model. Both strategies have their strengths, and the optimal choice depends on balancing complexity and performance to identify the most suitable approach.

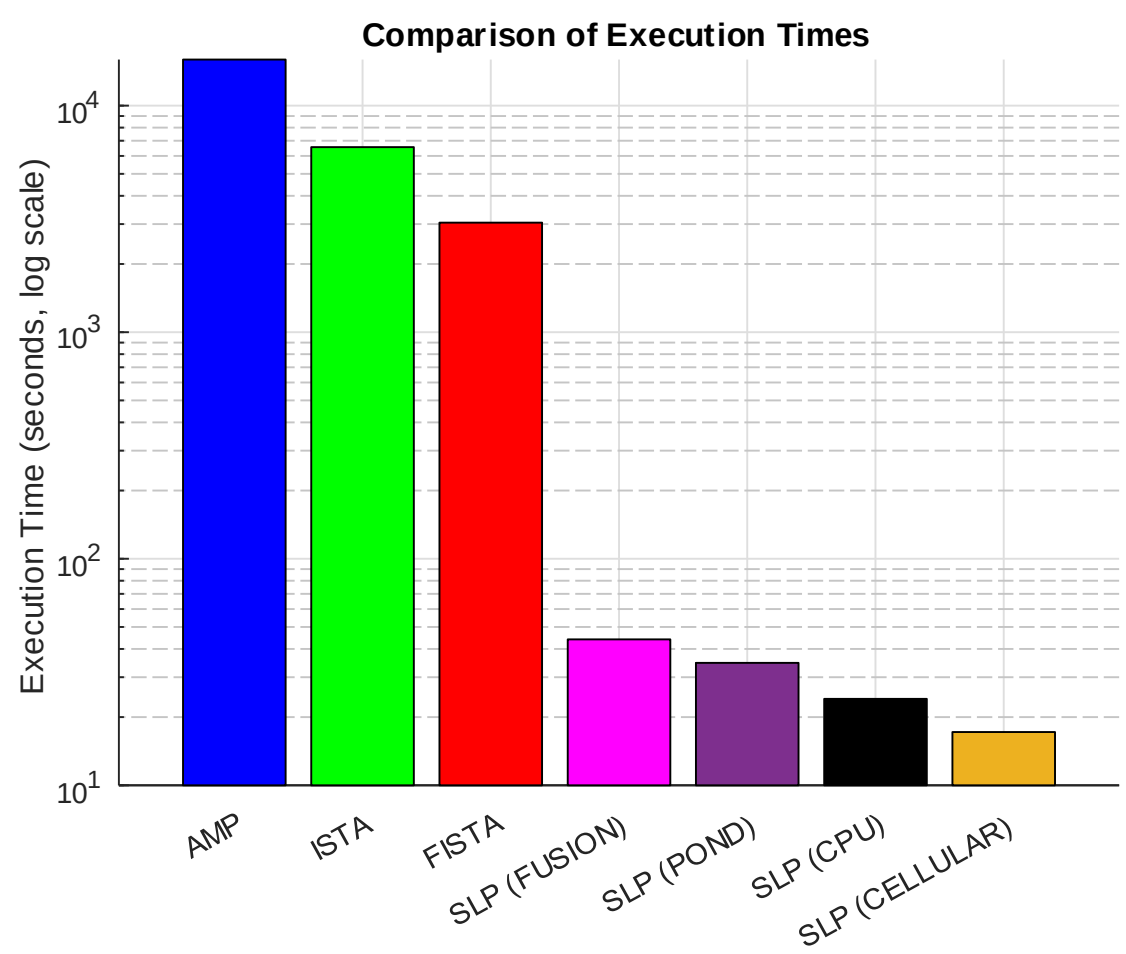

**FIGURE 11.** Comparison of execution times across all methods.

### *E. TIME EFFICIENCY AND COMPLEXITY OF THE DL AND BASELINE ALGORITHMS*

In addition to performance, a key criterion for evaluating an activity detection system is its response time, especially when deployed in real-world environments.The response time is crucial for practical applicability, as it directly affects the system's ability to perform in real-time applications. Fig. 11 compares the inference execution times of all algorithms previously evaluated. For simplicity, each curve in figures 8, 9, and 10 is represented with the same color scheme in this figure. The simulations were conducted on a high-performance GPU computing setup designed to redirect jobs to GPU resources. The simulations utilized an NVIDIA HGX A100 GPU (graphics processing unit) cluster, configured with 9 nodes, each equipped with 4 GPUs and a total of 10,120,000 MB of memory per node. This setup ensured consistent and efficient computation for all algorithms.

**TABLE 6.** Algorithm and MACs Complexity

| Algorithm | | MAC Operations |
|---|---|---|
| AMP | | $12tLKN$ |
| ISTA | | $8tLKN$ |
| FISTA | | $8tLKN$ |
| NN | DMLP (STRATEGY II) | $T[2NLV + (Z - 1)V^2 + KV]$ |
| | SLP (STRATEGY II) | $TV[2NL + K]$ |

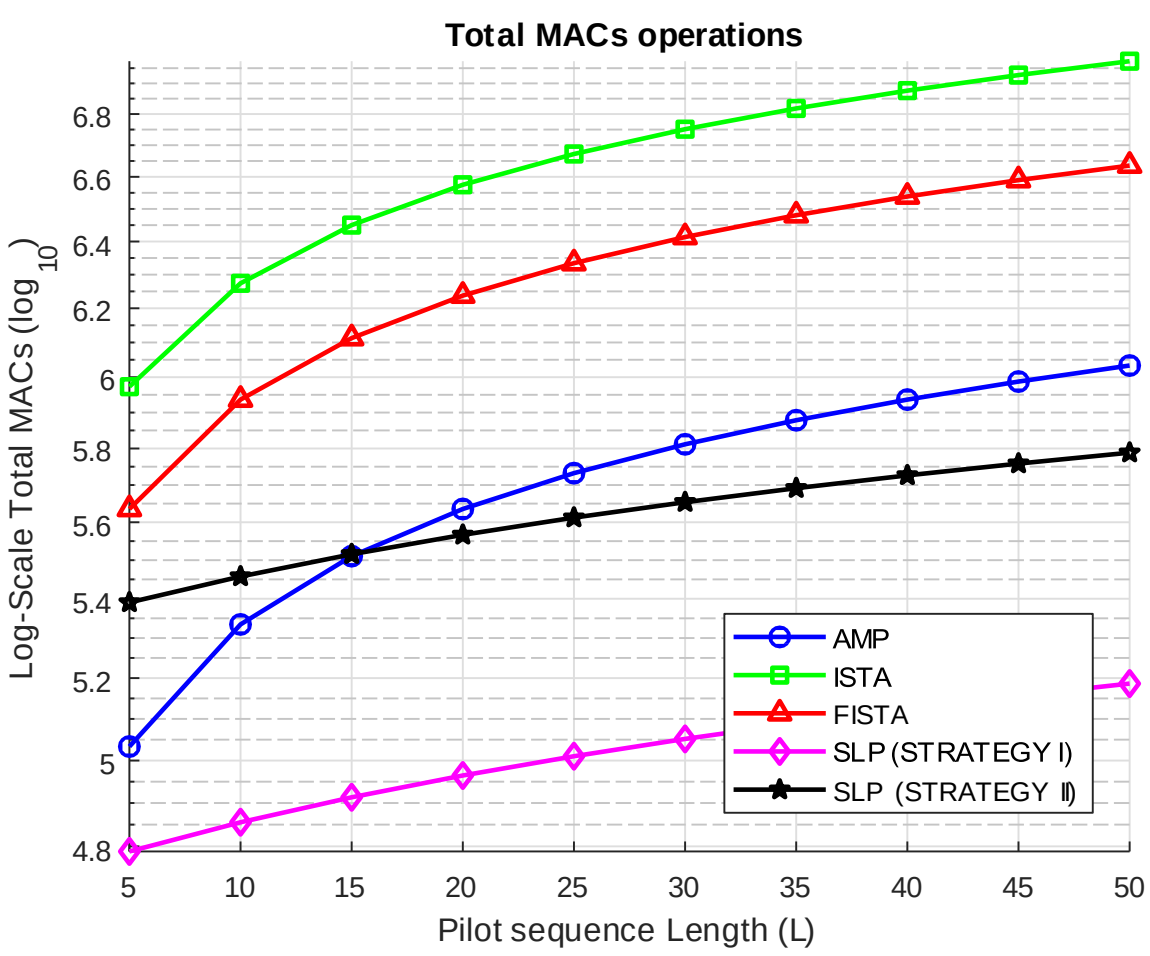

**FIGURE 12.** Computational complexity: MAC operations vs. pilot sequence length L.

For the iterative algorithms from the state-of-the-art [25], the number of iterations ($t$) is a critical parameter influencing convergence. In our experiments, we followed the iteration numbers specified in [25], setting $t = 18$ for AMP, $t = 100$ for FISTA, and $t = 235$ for ISTA. The comparison in Fig. 11, presented on a logarithmic scale, reveals the significant differences in execution times between data-driven and model-driven algorithms. Data-driven models demonstrate exceptionally low execution times compared to model-driven methods. However, as previously mentioned, the optimal choice depends on balancing complexity and performance.

For example, the decentralized approach with the ponderation strategy offers a favorable trade-off, achieving fast response times with minimal performance degradation. On the other hand, the centralized (CPU) approach, while slightly less effective in terms of performance, provides even faster response times and serves as a strong suboptimal detector. Although model-driven methods can occasionally achieve higher performance, this comes at the cost of considerably longer execution times, as shown in Fig. 11.

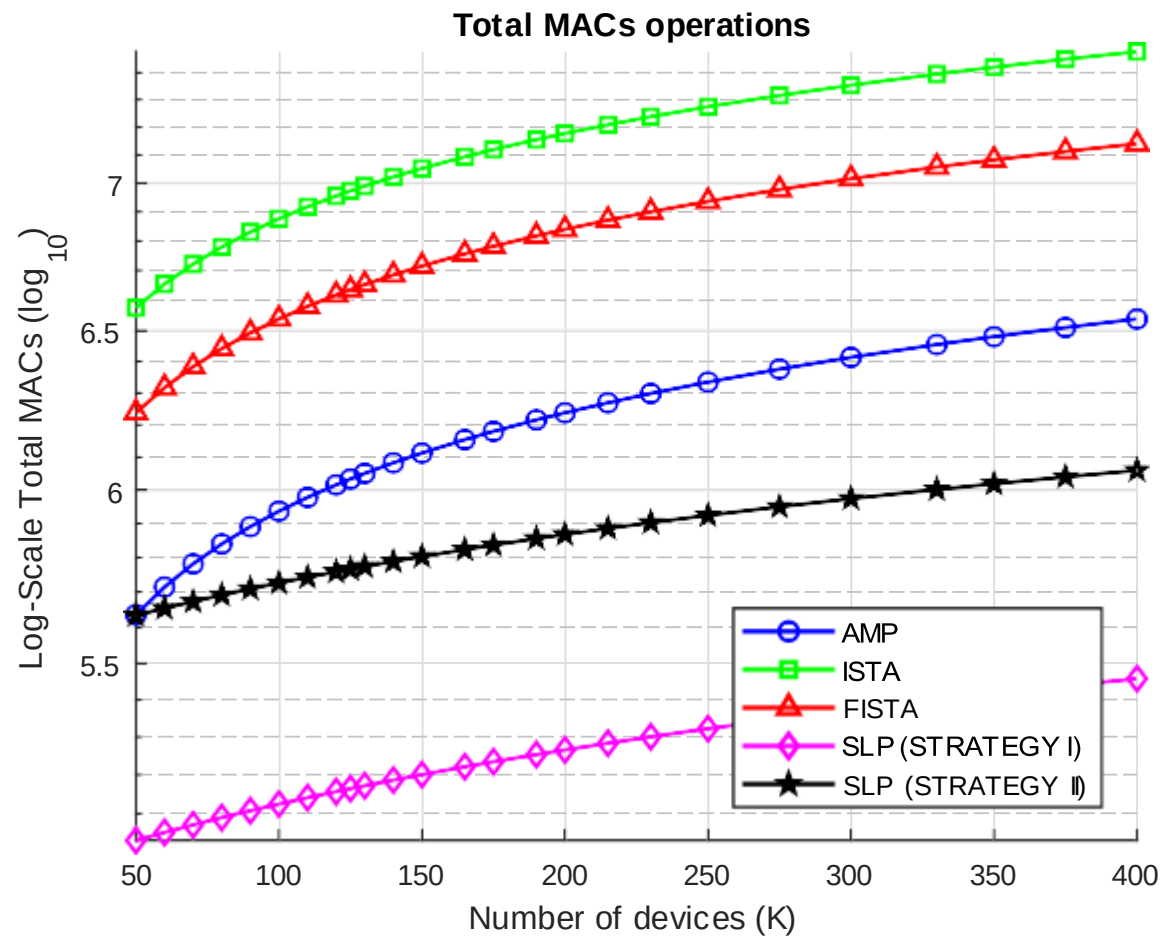

**FIGURE 13.** Computational complexity: MAC operations vs. number of devices.

Another key metric for evaluating an algorithm's efficiency is the number of multiply–accumulate (MAC) operations, which reflects its computational complexity. This measure helps assess the resources required for deployment, particularly in large-scale networks, highlighting the trade-off between performance and computational cost.

Table 6 outlines the computational complexity of various algorithms, expressed through the number of Multiply-Accumulate (MAC) operations. For the algorithms from literature such as AMP, ISTA, and FISTA, the formulas depend on the iteration count $t$, as previously defined. For the proposed architecture in centralized deployment (Strategy II), the general formula is provided, while the distributed deployment (Strategy I) corresponds to setting $T = 1$ in the same expression.

Focusing on the SLP architecture used in all simulations, the MAC counts listed in the table are equivalent to those of the deep multilayer perceptron (DMLP) architecture with $Z = 1$, reflecting a single hidden layer. This simplification aligns with the architectural constraints of the SLP design.

The results reveal distinct trends in complexity (Fig. 12, Fig. 13). Across varying $K$ and $L$, the distributed deployment (Strategy I) consistently exhibits the lowest computational complexity. The centralized deployment (Strategy II) also demonstrates significant efficiency, outperforming all conventional algorithms (AMP, ISTA, FISTA) in most scenarios. For fixed $K$, when $L$ is varied (Fig. 12), Strategy II ranks second in complexity after Strategy I. However, at very low values of $L$, AMP occasionally shows marginally lower complexity than Strategy II. This advantage is limited to small $L$, as AMP's performance diverges under such conditions (Fig. 8), creating a trade-off between complexity and reliability. Notably, ISTA and FISTA consistently exhibit the highest complexity across all configurations. For fixed $L$, when $K$ is varied (Fig. 13), the complexity hierarchy remains unchanged: Strategy I (lowest), Strategy II, AMP, FISTA, and ISTA (highest).

This study provides two complementary comparative analysis, focusing on two key metrics: execution time (measured under uniform hardware conditions) and computational complexity in terms of MAC operations. The results demonstrate the superiority of the proposed model over all existing literature-based algorithms, positioning it as an optimal detector that achieves robust detection probability with significantly lower complexity and faster response time.

## VII. CONCLUSION

This paper explored device activity detection in GF-RA scenarios within CF-mMIMO networks, introducing centralized and decentralized approaches built on a tailored deep learning framework. The findings highlighted the advantages of CF networks over cellular systems, including improved connectivity and reduced outage probabilities. Additionally, the analysis of execution time as well as computational complexity underscored the efficiency of data-driven algorithms compared to iterative, model-based methods. The proposed approaches demonstrated excellent performance in detecting active devices while ensuring low computational complexity and fast inference execution times, making them well-suited for real-world mMTC deployments.

## ACKNOWLEDGMENT

This work was supported from the French National Research Agency (ANR-22-CE25-0015) within the frame of the project POSEIDON.

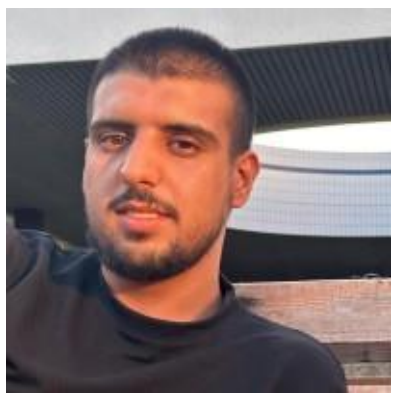

**ALI ELKESHAWY** received the diploma in telecommunication, communication, and computer engineering from the Lebanese University, Faculty of Engineering III (ULFG), Beirut, Lebanon, in 2021, and the M.Sc. degree in telecommunication, networks, and security from Saint Joseph University (USJ-ESIB), Mansouriyeh, Lebanon, in 2022.

He is currently pursuing the Ph.D. degree at CentraleSupélec, Rennes, France, as a member of the SIGNAL research group of the CNRS Institute of Electronics and Digital Technologies (IETR). His research focuses on artificial intelligence applications in telecommunications, particularly physical layer techniques for massive machine-type communications and grant-free random access in cell-free massive MIMO systems. His expertise includes AI-driven wireless communications for IoT intrusion detection, natural language processing for speech recognition and machine translation, as well as embedded system deployment and practical network design.

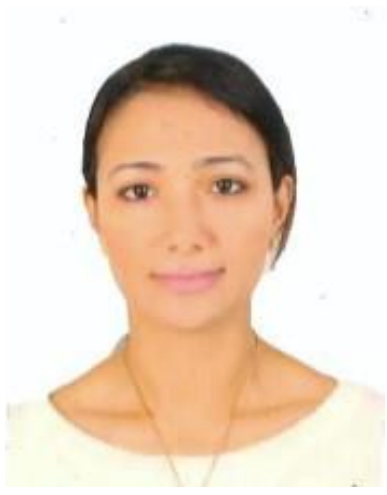

**HAÏFA FARÈS** received bachelor's and M.Sc. degrees in telecommunication engineering from the Higher School of Communications of Tunis (Sup'Com), in 2007 and 2008, respectively, and the Ph.D. degree in digital communications from IMT-Atlantique, France, in 2011.

She is currently an Associate Professor at CentraleSupélec of Rennes. She is a member of the SIGNAL research group of the CNRS Institute of Electronics and Digital Technologies (IETR), Rennes Laboratory. Her research interests include the area of communication theory, including nonlinear modulations, indexed modulations, energy-efficient communications, non-orthogonal multiple access, physical layer security and integration of machine learning into future communication systems.

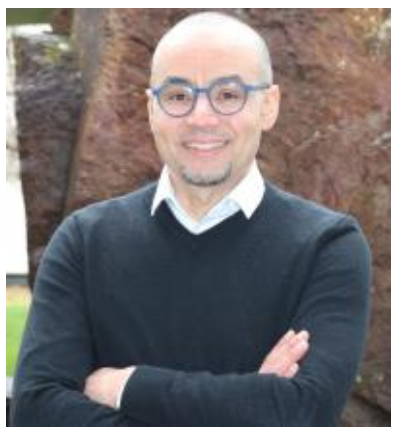

**AMOR NAFKHA** (Senior Member 2016, IEEE) received the B.Sc. degree in information and communications technology engineering from the Higher School of Communications (Sup'Com), Tunis, Tunisia, in 2001, and the Ph.D. degree in information and communications technology from the University of South Brittany (UBS), Lorient, France, in 2006. From 2006 to 2007, he was a Postdoctoral Researcher with the Signal, Communication, and Embedded Electronics (SCEE-IETR) Research Group, CentraleSupélec, France, where he has also been an Associate Professor since January 2008. In November 2021, he obtained the Habilitation à Diriger des Recherches from the University of Rennes 1 (UR1), Rennes, France, and he is promoted to full professor at CentraleSupelec in 2024.

He was actively involved in the reconfigurable hardware platform implementation for software-defined radio and coauthoring several contributions on FPGA dynamic partial reconfiguration. He has published more than 80 papers in international peerreviewed journals and conferences. His research interests include multiuser and MIMO detection, hardware implementation, information theory, sample rate conversion, and spectrum sensing techniques. His current research interests include physical layer security in wireless communications, radio frequency fingerprinting, and embedded AI security.